\documentclass[12pt,epsfig]{article}
\usepackage{axodraw}
\usepackage{graphicx}
\usepackage{epsfig}

\newcommand{\slsh}[1]{\not{\hbox{\kern-2pt${#1}$}}}

\newcommand{\ba}[1]{\begin{eqnarray} \label{#1}}
\newcommand{\ea}{\end{eqnarray}}

\def\beq{\begin{equation}}
\def\eeq{\end{equation}}
\def\bea{\begin{eqnarray}}
\def\eea{\end{eqnarray}}
\def\bqu{\begin{quote}}
\def\equ{\end{quote}}

\def\gappeq{\mathrel{\rlap {\raise.5ex\hbox{$>$}}
{\lower.5ex\hbox{$\sim$}}}}

\def\lappeq{\mathrel{\rlap{\raise.5ex\hbox{$<$}}
{\lower.5ex\hbox{$\sim$}}}}

\begin{document}
\pagestyle{empty}

\begin{flushright}
{CERN-PH-TH/146, KCL-PH-TH/2011-21, LCTS/2011-04}\\
{UHU-GEM/29}\\
\end{flushright}

\vspace*{1 cm}
\begin{center}
{\large {\bf Searches for Lepton Flavour Violation at a Linear Collider}} \\
\vspace*{1cm}
{\bf E. Carqu\'{\i}n$^1$, J.~Ellis$^{2,3}$, M.E. G{\'o}mez$^4$, S.~Lola$^{2,5}$
}\\
\vspace{0.3cm}
$^1$ Centre of Subatomic Studies, Technical University Federico Santa Mar\'ia,
Valpara\'iso, Chile \\
$^2$ Theory Division, Physics Department, CERN, CH-1211 Geneva 23, Switzerland \\
$^3$ Theoretical Particle Physics and Cosmology Group, Department of Physics, 
King's College London, Strand, London WC2R 2LS, UK \\
$^4$ Department of Applied Physics, University of Huelva, 21071 Huelva, Spain \\
$^5$ Department of Physics, University of Patras, 26500 Patras, Greece 

\vspace*{2cm}
{\bf ABSTRACT}  
\end{center}
\noindent
We investigate the prospects for detection of lepton flavour violation in sparticle production and decays at a Linear Collider (LC), in models guided by neutrino oscillation data. We consider both slepton pair production and sleptons arising from the cascade decays of non-leptonic sparticles. We study the expected signals when lepton-flavour-violating (LFV) interactions are induced by renormalization effects in the Constrained Minimal Supersymmetric extension of the Standard Model (CMSSM), focusing on the subset of the supersymmetric parameter space that also leads to cosmologically interesting values of the relic neutralino LSP density. Emphasis is given to the complementarity between the LC, which is sensitive to mixing in both the left and right slepton sectors, and the LHC, which is sensitive primarily to mixing in the right sector. We also emphasize the complementarity between searches for rare LFV processes at the LC and in low-energy experiments.

\noindent

\setcounter{page}{1}

\pagestyle{plain}
\newpage
\section{Introduction}

In recent years, a plethora of data from atmospheric~\cite{skatm}, solar~\cite{sksol} and
long-baseline reactor~\cite{KamLand} and accelerator~\cite{K2K,MINOS}
neutrino experiments have established the existence of neutrino masses and
oscillations with near-maximal  $\nu_\mu - \nu_\tau$
and large $\nu_e \to \nu_{\mu}$ mixing. 
A natural expectation in this context is that
charged-lepton-flavour violation (LFV) should occur at some level.
This may be enhanced sufficiently to become observable in a class of theories predicting new physics
at the TeV scale accessible to colliders, particularly in supersymmetric theories.
There are several sources of lepton flavour
violation in such theories, which could have
unacceptably large LFV if the soft supersymmetry-breaking masses of different
sfermion flavours were not universal at some level. For this reason, it is often
assumed that these masses are equal at the grand-unification scale, as in
the Constrained Minimal Supersymmetric extension of the Standard Model
(CMSSM).

Even if the sfermion mass matrices are 
diagonal at the unification scale, as in the CMSSM, quantum corrections would modify this structure
while running from the GUT scale to low energies. This effect is particularly interesting
in see-saw models for neutrino masses, where the Dirac neutrino Yukawa couplings
cannot be diagonalised simultaneously with the charged-lepton and slepton mass matrices~\cite{bm}.
Given the large mixing of the corresponding neutrino species,
charged-lepton-flavour violation may occur at enhanced rates for sufficiently 
small soft supersymmetry-breaking masses,
giving rise to observable signals such as $\mu \to e \gamma$, $\mu-e$ conversion,
$\tau \to \mu \gamma$ and $\tau \to e
\gamma$ ~\cite{rev,LFVhisano, LFVres}. 

Other charged-lepton-flavour 
violating possibilities that have been considered include 
slepton pair production at a Linear Collider (LC)~\cite{sleptonscemu,LFV-LC,
LC21,LC2,LFV-LC2}, and also signals 
 at the LHC \cite{LHC1,LHC2,CEGLR2}, particularly in
$\chi_2^0\to \chi + e^\pm \mu^\mp$ 
$\chi_2^0\to \chi + \mu^\pm \tau^\mp$ decays 
(here $\chi$ is the lightest
neutralino, assumed to be the lightest supersymmetric particle (LSP),
and $\chi_2^0$ is the second-lightest neutralino).
These decays potentially provide search prospects
that are complementary to direct searches for the flavour-violating decays
of charged leptons. However,
most of the previous studies of LFV at a LC use a low-mass
spectrum (focusing on $P_{cm}<400{\rm ~GeV}$ or $\sqrt{s}<800{\rm ~GeV}$) and the points chosen 
for study do not always
satisfy all the relevant
phenomenological and experimental bounds (e.g.,~\cite{LEPH}), 
which are now being improved by the LHC~\cite{LHC,MC6}. Moreover, 
previous analyses have not always taken into account the constraints on the cosmological
relic LSP density imposed by WMAP and other experiments~\cite{WMAP}.

In the current paper, we revisit the various detection channels for
lepton flavour violation at a Linear Collider, studying the
complementarity between different processes. In all cases,
we focus on regions that satisfy not only all phenomenological
constraints, but also the cosmological relic density considerations. 
We extend previous results to somewhat heavier sparticle spectra,
and make comparisons with the expectations for experimental sensitivity
at the LHC.

Our paper is structured as follows. In Section 2,  we briefly review the theoretical framework and
sources of slepton mixing and summarize 
the possibilities for observable signatures
in slepton production at a LC.
In Section 3, we discuss 
the relevant supersymmetric parameter space,  and pick
representative points motivated by both
phenomenological and cosmological considerations. 
In Section 4, we look at the expected slepton
mixing parameters, with emphasis on the region where LC experiments could give
information additional to that obtainable from low-energy LFV decays.
In Section 5, we discuss in more detail the
cross sections, comparing with the expected signals at the LHC, where
appropriate.
Finally, in Section 6, we summarize our conclusions and outlook.

\section{LFV in Slepton Pair Production}

\subsection{CMSSM with LFV}

In the unrotated charged-lepton flavour basis $\tilde{\ell}_i=\left( \tilde{e}_L,\tilde{\mu}_L,\tilde{\tau}_L,
\tilde{e}_R^*,\tilde{\mu}_R^*,\tilde{\tau}_R^* \right)$, the charged slepton mass matrix is:
\beq
M_{\tilde{\ell}}^2= \left(\begin{array}{cc}
M_{LL}^2&M_{LR}^2\\
M_{RL}^2&M_{RR}^2\\
\end{array}
\right) ,
\eeq
where
\bea\label{defsSofts}
M^2_{LL}&=&m_\ell^\dagger m_\ell+ M_L^2-\frac{1}{2}(2 m_W^2- m_Z^2) 
\cos{2\beta}  \ {I} , \nonumber \\
M^2_{RR}&=&m_\ell^\dagger m_\ell + M_R^2-( m_Z^2- m_W^2) \cos{2\beta} \ {I} ,
\nonumber \\
M^2_{LR}&=& \left( A^e- \mu \tan{\beta} \right) \ m_\ell , \nonumber \\
M^2_{RL}&=&(M^2_{LR})^\dagger .
\eea
Here we parametrize trilinear soft supersymmetry-breaking terms as
$A^e_{ij} (\lambda_e)_{ij}$, where the $\lambda_e$ are the respective
Yukawa couplings.
For universal soft terms at some high input scale, one has
\bea
M_L^2 = M_R^2 \ = \ m_0^2 \ {I} ,  ~~ A^e_{ij}=A_0\delta_{ij} ,
\eea
whereas flavour-mixing entries may be parametrized by:
\beq
\delta_{XX}^{ij}=(M^2_{XX})^{ij}/(M^2_{XX})^{ii}\;\;\;\;\;\;(X=L,R).
\eeq
The correspondence between the mixing parameters $\delta_{XX}$ 
and the flavour mixing parameters used in the phenomenological
study of slepton production cross sections 
is easily derived. For example, defining the splitting 
of the third-generation soft supersymmetry-breaking mass terms as
might be generated by the renormalization-group equations (RGEs) by
\beq
K_{XX}=1-(M^2_{XX})^{33}/(M^2_{XX})^{22} ,
\eeq
the slepton mixing angle $\tilde{\theta}_{23}$ 
and the  splitting of the second- and third-generation
mass eigenstates $\Delta \tilde{m}_{23}$ can be written as:
\bea
\tan(2\tilde{\theta}_{23})&=&2 \delta_{XX}/K_{XX} , \\
\Delta \tilde{m}_{23}&=&\frac{(M^2_{XX})^{22}}{\tilde{m}} \frac{\delta_{XX}}
{\sin(2\tilde{\theta}_{23})}
\eea
where $\tilde{m} = \frac{1}{2}(m_2+m_3)$.

The evaluation of the LFV observables is done by performing
the diagonalization of the slepton mass matrices 
(see~\cite{LFVhisano}, for instance), inserting the full  
rotation matrices in the lepton-slepton-gaugino vertices and summing over 
all the mass eigenstates of the exchanged particles.

\subsection{LFV Cross Sections}

As already discussed in the introduction, charged-lepton flavour violation
at a LC may occur either directly in slepton pair production or indirectly via slepton
production in cascade decays~\cite{sleptonscemu}. Processes leading
to lepton production in the decays of a pair of sleptons include
\bea
e^+ e^- & \rightarrow & \tilde{\ell}_i^{-} \tilde{\ell}_j^{+} \rightarrow 
\tau^{\pm} \mu^{\mp} \tilde{\chi}^{0}_1 \tilde{\chi}^{0}_1 , \nonumber \\
e^+ e^- & \rightarrow & \tilde{\nu}_i \tilde{\nu}_j^c  \rightarrow 
\tau^{\pm} \mu^{\mp} \tilde{\chi}^{+}_1 \tilde{\chi}^{-}_1, 
\label{eq:pair}
\eea
for which representative Feynman diagrams are shown in Fig.~1. 
Slepton production may also result from the
the cascade decays of the heavier gauginos, e.g., via the processes
\bea
e^+ e^- & \rightarrow & \tilde{\chi}_2^{\pm} \tilde{\chi}_1^{\mp} \rightarrow
\tau^{\pm} \mu^{\mp} \tilde{\chi}^{+}_1 \tilde{\chi}^{-}_1 , \nonumber \\
e^+ e^- & \rightarrow & \tilde{\chi}_2^0 \tilde{\chi}_1^0 \rightarrow 
\tau^{\pm} \mu^{\mp} \tilde{\chi}^{0}_1 \tilde{\chi}^{0}_1 .
\label{eq:single}
\eea
These amplitudes differ in the interference terms,
and read as follows:
\begin{eqnarray}
&&{ M}^{\rm pair}_{\alpha\beta}= \sum_i { M}^{\rm pair}_P  \frac{i}
{q^2-\tilde{m}^2_i+i\tilde{m}_i\Gamma_i} T_{i\alpha} { M}^+_D  \frac{i}
{p^2-\tilde{m}^2_i+i\tilde{m}_i\Gamma_i} T^*_{i\beta} { M}^-_D ,
\label{pairLFV}
\\
&&{ M}^{\rm casc}_{\alpha\beta}= \sum_i { M}^{\rm casc}_P T_{i\alpha} \frac{i}
{q^2-\tilde{m}^2_i+i\tilde{m}_i\Gamma_i} T^*_{i\beta} { M}^-_D ,
\label{singleLFV}
\end{eqnarray}
for pair production and cascade slepton production, respectively,
where ${ M}_P$ and ${ M}_D$ are the production   
and decay
amplitudes for sleptons in the absence of LFV, and 
$T_{i\alpha}$ parametrizes the lepton-flavour-mixing matrix element.

\begin{figure}[!t]
\begin{center}
\begin{picture}(235,50)(0,48)
                           \ArrowLine(50,50)(10,90)        \Text(12,100)[r]{$e^{+}$}
                           \ArrowLine(10,10)(50,50)        \Text(12,0)[r]{$e^{-}$}              \Vertex(50,50){2}
                           \Photon(50,50)(100,50){3}{5}           \Text(75,54)[b]{\(\gamma,Z\)} \Vertex(100,50){2}
                           \DashArrowLine(130,80)(100,50){5}  \Text(108,70)[b]{\(\tilde{\ell}^{+}_{j}\)}              \Vertex(130,80){2}
                           \ArrowLine(160,100)(130,80)        \Text(162,100)[l]{\(\ell_{\beta}^{+}\)}
                           \ArrowLine(130,80)(160,60)        \Text(162,60)[l]{\(\tilde{\chi}^{0}_{b}\)}
                           \DashArrowLine(100,50)(130,20){5}  \Text(108,30)[t]{\(\tilde{\ell}_{i}^{-}\)}          \Vertex(130,20){2} 
                           \ArrowLine(130,20)(160,40)        \Text(162,40)[l]{\(\ell^{-}_{\alpha}\)}
                           \ArrowLine(160,0)(130,20)        \Text(162,0)[l]{\(\tilde{\chi}_{a}^{0}\)}
                         \end{picture}
                         \begin{picture}(100,50)(5,48)
                           \ArrowLine(20,80)(0,100)   \Text(0,105)[r]{$e^{+}$}
                           \ArrowLine(0,0)(20,20)     \Text(0,-5)[r]{$e^{-}$} 
                           \ArrowLine(20,20)(20,80)   \Text(22,50)[l]{\(\tilde\chi_c^{0}\)} \Vertex(20,20){2} \Vertex(20,80){2}
                           \DashArrowLine(40,100)(20,80){5} \Text(28,93.5)[b]{\(\tilde{\ell}_{j}^{+}\)} \Vertex(40,100){2}
                           \ArrowLine(70,100)(40,100)  \Text(72,100)[l]{\(\ell_{\beta}^{+}\)}
                           \ArrowLine(40,100)(70,70) \Text(70,70)[l]{\(\tilde{\chi}_{b}^{0}\)}
                           \DashArrowLine(20,20)(40,0){5} \Text(28,5)[t]{\(\tilde{\ell}^{-}_{i}\)}    \Vertex(40,0){2}
                           \ArrowLine(40,0)(70,30)  \Text(70,0)[l]{\(\tilde{\chi}_{a}^{0}\)}
                           \ArrowLine(70,0)(40,0) \Text(70,30)[l]{\(\ell^{-}_{\alpha}\)}
\end{picture}

\vspace{2cm}
\noindent
\caption{\small \it Feynman diagrams for 
$e^+e^-\to \tilde{\ell}_j^+\tilde{\ell}^-_i 
\to \ell_\beta^+
\ell^-_\alpha\tilde{\chi}^0_b\tilde{\chi}^0_a$.
The arrows on scalar lines indicate the flow of lepton number.
Similar diagrams - appropriately modified - exist for charginos.}
\label{figs}
\end{center}
\end{figure}
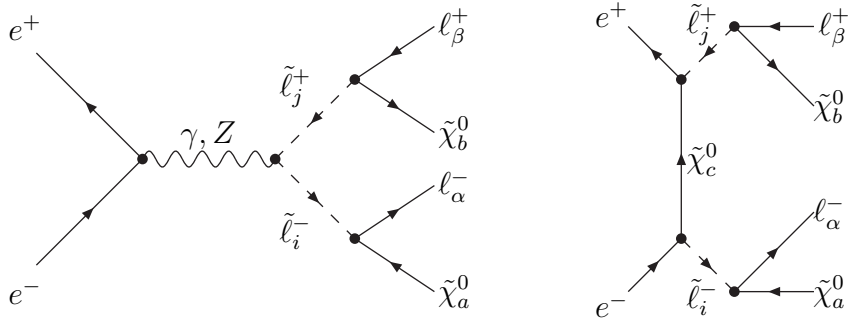

Complete expressions for the respective cross sections
are given in~\cite{LFV-LC}, and used
in our work. However, intuition may be developed by first looking at the simplified case
including only 2-3 slepton mixing. In this case, the 
only diagrams 
contributing to the cross sections are those mediated by
$\gamma$ and $Z^0$ exchange. Even simpler formulae can 
be obtained in certain limits, such as the 
cases of narrow widths and small mass differences between the
sleptons of different generations.
For example, if
$\Delta \tilde{m}_{ij} \ll \tilde{m} = \frac{1}{2}(m_2+m_3)$ 
and
$\tilde{m}\overline{\Gamma}_{ij} 
\simeq (\tilde{m}_i\Gamma_i+\tilde{m}_j\Gamma_j)/2\ll
\tilde{m}^2$,  
in the presence of 
2-3  mixing the cross sections for pair and cascade
slepton production are given by~\cite{JK-APP}:     
\begin{eqnarray}
&&\sigma^{\rm pair} = \chi_{23}(3-4 \chi_{23})  
\sin^2 2\tilde{\theta}_{23} \;
\sigma(\bar f\,f\to \tilde{\ell}^+ \,
\tilde{\ell}^-) 
Br(\tilde{\ell}^+ \to \ell^+ \, X)  
Br(\tilde{\ell}^- \to  \ell^-\, Y) , \label{crosspair}
\label{eq:kali1}\\[2mm]
&&\sigma^{\rm casc}_{\alpha\beta}=\chi_{23} \sin^2 2\tilde{\theta}_{23}\;
\sigma(f\,f'\to \ell^+ \, X\, \tilde{\ell}^-)
Br( \tilde{\ell}^- \to \ell^-\, Y) ,
\label{eq:kali2}
\end{eqnarray}

where $X, Y$ represent gauginos in the final state, and 
$\sigma(\bar f\,f\to \tilde{\ell}^+ \,
\tilde{\ell}^-)$ and $Br( \tilde{\ell}^\pm \to \ell^\pm X)$ 
are the cross sections 
and branching ratios in the absence of flavour violation. In the case of 
flavour conserving cascade decays, 
$\sigma(f\,f'\to \ell^+ \, X\, \tilde{\ell}^-)$ includes  
the intermediate production of two gauginos and the subsequent decay of the heavier into 
a  lepton-slepton pair. 
LFV enters via $\chi_{23} \equiv x_{23}^2/2(1+x_{23}^2)$
where $x_{23} \equiv \Delta \tilde{m}_{23}/\overline{\Gamma}_{23}$, and
$\sin^2 2\tilde{\theta}_{23}$ parametrizes the slepton mixing angle. In the 
limit $x_{23}\gg 1$, interference can be neglected and the results are even 
further simplified. However, in the case that the 
interference term dominates, it suppresses LFV significantly.

The exact experimental signatures of the above effects depend on the decay
chains, as discussed in previous analyses \cite{LFV-LC}. Channels 
involving the lightest charginos are potentially interesting, particularly those
with hadronic chargino decays, since the lightest
neutralino is `seen' as missing energy.
In general, the dominant chargino decay is $\tilde{\chi}_1^+\rightarrow 
\tilde{l} \nu$, whereas channels such as $\tilde{\chi}_1^+\rightarrow q_u q_d  
\tilde{\chi}^0$ are relevant  when $m_{\tilde{\chi}_1^+}<m_{\tilde{\tau}_1}$ 
or, in other words, when the ${\tilde{\chi}_1^+}$ is the NLSP.
Channels leading to hadronic  $\tilde{\chi}_1^+$ decays~\cite{LFV-LC}
are mediated only by left sleptons, whilst
$e^+e^-\rightarrow \tau ^\pm\mu^\mp +2 \chi^0$
may be mediated by either right or left sleptons.  However,
points with  $m_{\tilde{\chi}_1^+}<m_{\tilde{\tau}_1}$ are disfavored
in the framework of the CMSSM. For instance, the previously proposed 
point with $\tan\beta=3$, $m_0=100$ ${\rm ~GeV}$, $M_{1/2}=200$ ${\rm ~GeV}$~\cite{LC21}
was excluded by the LEP $m_h$ bound~\cite{LEPH} and is now incompatible also with LHC limits~\cite{LHC}.
This could perhaps be achieved in the focus-point 
region, where $M_{1/2} <$ 1 TeV
and $m_0>$  several TeV, but slepton masses in this region are
very heavy and hence LFV processes are more suppressed.

The channel $e^+e^-\rightarrow \tau ^\pm\mu^\mp +2 \chi^0$ was studied in~\cite{LFV-LC2}, where
it was found that a signal of $\sim~1$~fb could be distinguished from the 
background. Only slepton mixing in the left sector arising from 
see-saw neutrino mass models was included in~\cite{LFV-LC2},  and the limit 
from  $\tau\rightarrow \mu \gamma$ was at the time less restrictive.
As shown in~\cite{CEGLR2}, reasonable rates of LFV processes at the LHC 
in the slepton sector are possible while preserving the bound on 
$\tau\rightarrow \mu \gamma$ if one assumes mixing also among the right 
sleptons, as may occur in non-minimal GUT theories.

\section{Supersymmetric Parameter Space}
\label{SUSY-space}

From the above discussion, it is clear that  the dependence on  the
supersymmetric parameter space is particularly crucial, since
the chains of cascade decays that will dominate,
as well as the relative strengths of the signals and backgrounds,
depend on the mass hierarchies of the superpartners.

We start with some preliminary considerations of the relevant channels
within the CMSSM parameter space.
In previous work \cite{CEGLR2,pedro} a region that is promising
from the cosmological point of view was discussed, in which
sizable $\tilde{\tau}-\chi$ coannihilations or direct-channel $H/A$ resonances
 lead to values of
$\Omega_{\chi}h^2$ consistent with WMAP. 
As a representative of the coannihilation case, we chose $\tan\beta=35$, a value that
leads to a relatively heavy sparticle spectrum compatible with WMAP.
At $\tan\beta=45$ we found points in the resonance funnel, 
which have larger scalar masses since $m_0$ can take higher values than in the coannihilation area. We did not consider the focus-point region 
of the CMSSM because the  super-heavy slepton masses there lead to reduced LFV effects.
We note that these regions are compatible with the LHC constraints on supersymmetry,
which indicate that $M_{1/2} > 450$~GeV and favour $\tan \beta > 15$~\cite{MC6}.

\begin{figure}[!t]
\begin{center}
\includegraphics[scale=.40]{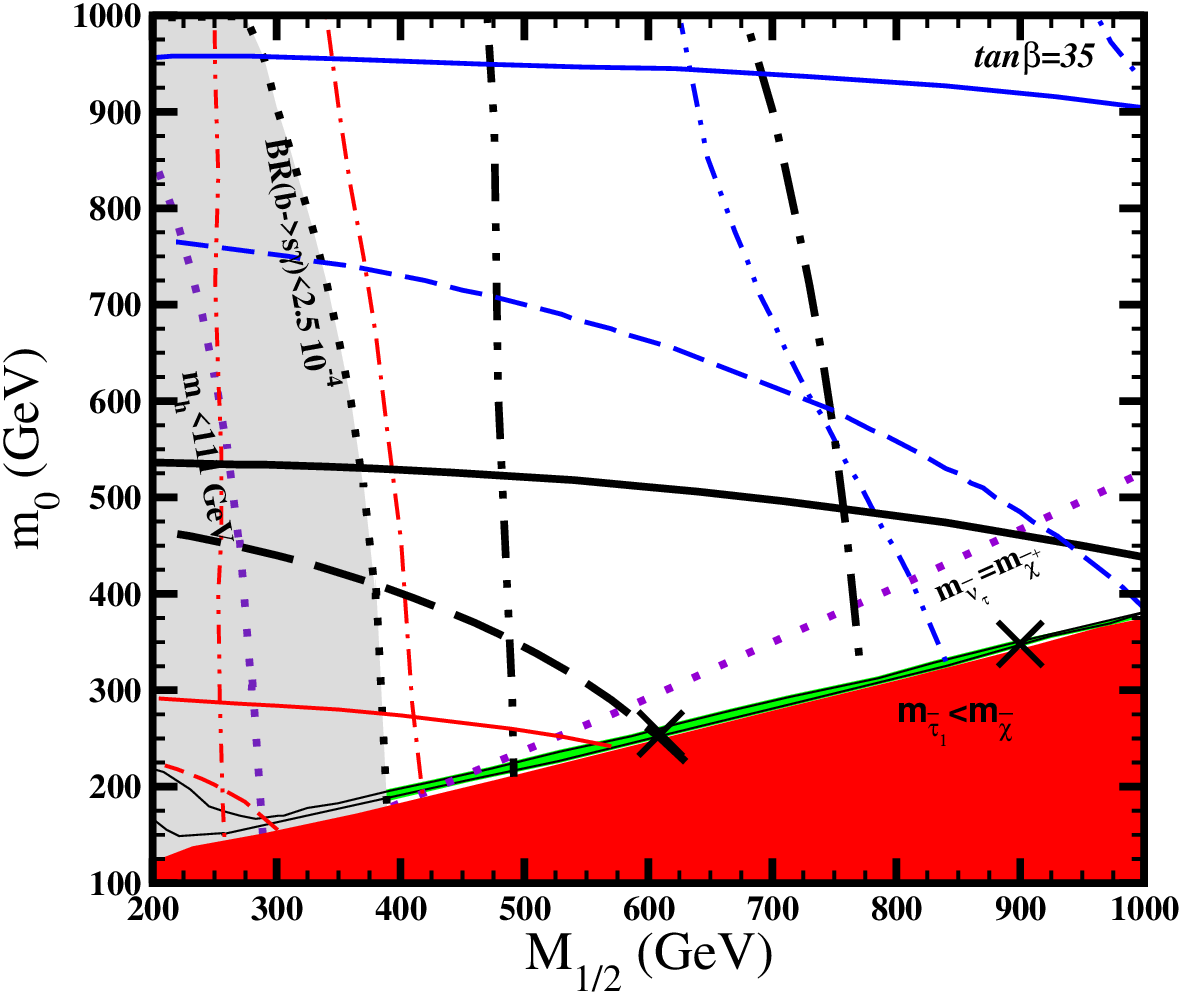} 
\includegraphics[scale=.40]{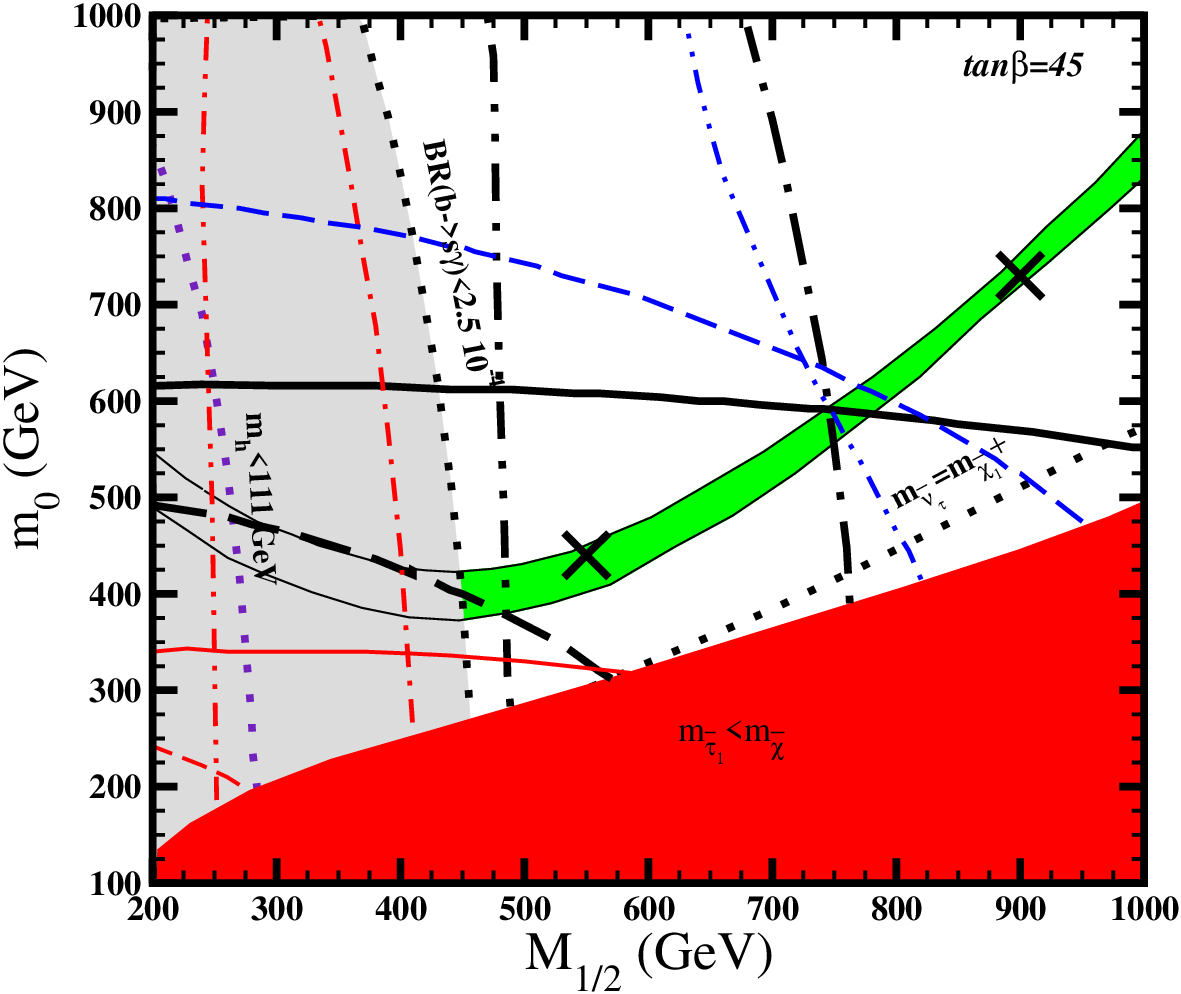} 
\end{center}
\caption
{\small \it Contours of the cross sections 
$\sigma (e^+e^-\rightarrow \tilde{l}^+\tilde{l}^-) = 1$~fb in the $(M_{1/2}, m_{0})$ planes
for $\tan\beta=35$ (left) and $45$ (right) at three different
energies, namely $\sqrt{s}=500$~GeV (red thin lines),  
$\sqrt{s}=1000$~GeV (black thick lines) and 
$\sqrt{s}=2000$~GeV (blue thin lines). The upper and lower near-horizontal lines correspond to the
following slepton pair production processes, respectively: $\tilde{\tau}_1^+\tilde{\tau}_1^-$ (solid) and 
$\tilde{\nu}_{\tau}\tilde{\nu}_{\tau}$ (dash). From left to right, the near-vertical lines correspond to
$\sigma (e^+e^-\rightarrow \chi_1^0\chi_2^0)=1$~fb (dot-dash) 
and $\sigma (e^+e^-\rightarrow \chi_1^+\chi_2^-) =1$~fb (double-dot-dash).
In regions above the line where $m_{\tilde{\chi}_1^+}=m_{\tilde{\nu}_\tau}$ (purple dots),
chargino-mediated processes are allowed. Representative contours of
BR($b\rightarrow s \gamma $) are indicated, and the regions with LSP relic density in
the range preferred by WMAP are shaded green. The benchmark points selected for
further study are indicated by crosses.
}
\label{fig:areaCS}
\end{figure}

The physical values of the masses are obtained by integrating the 
renormalization group equations from $M_{GUT}$ down to low energies. We use 
the two-loop supersymmetric renormalization group equations~\cite{Martin:1993zk} except for the trilinear terms, the gaugino and 
the sfermion masses, which are calculated at the one-loop level.  Electroweak 
Symmetry breaking (EWSB) occurs at the scale 
$M_{SUSY} = \sqrt{\tilde{m}_{t_1}\tilde{m}_{t_2}}$ (where the one-loop 
contributions to the scalar potential are less relevant). At this scale 
we incorporate the SUSY threshold corrections to $m_b$ , $m_\tau$ and $m_{t}$ 
by redefining the corresponding Yukawa couplings as done in ref. \cite{GNIS1}. 
From $M_{SUSY}$ to $M_Z$ we use the Standard Model renormalisation group 
equations. The running top mass is calculated iteratively, removing the Yukawa coupling $\lambda_t$, along with 
its derivative, from the remaining running from $m_t$ to $M_Z$. 

Within this framework, for every set of input parameters, we perform an iterative integration 
of the RGE's (back and forth, from high to low energies). We define 
$M_{GUT}$ as the meeting point of the gauge 
couplings $\alpha_1$  and $\alpha_2$ and use this 
scale to find the unified value of the 
couplings, $\alpha_U$ . The value $\alpha_3 (M_{GUT})$ is 
obtained from $\alpha_3 (M_{M_Z})$ (its deviation 
from $\alpha_U$ being very small). The third generation Yukawa couplings 
are obtained by using the following set of values: $m_\tau(M_Z) = 1.7463$~GeV 
(which takes into 
account the SM radiative corrections) and the top pole mass
$m_t = 172.6$~GeV \cite{mtop}. For $m_b$ we 
take $m_b (M_Z ) = 2.92$ GeV which is the value where $m(m_b) = 4.25$ GeV 
is mapped, taking $\alpha_s (MZ ) = 0.1172$. Note that the value of the bottom 
Yukawa coupling extracted from $m_b(M_Z)$ is very sensitive to the SUSY threshold 
corrections, especially at large $\tan\beta$; in our case we implement a 
complete one loop contributions as in ref.\cite{GNIS0}.

\begin{figure}[!h]
\begin{center}
\includegraphics[scale=.40]{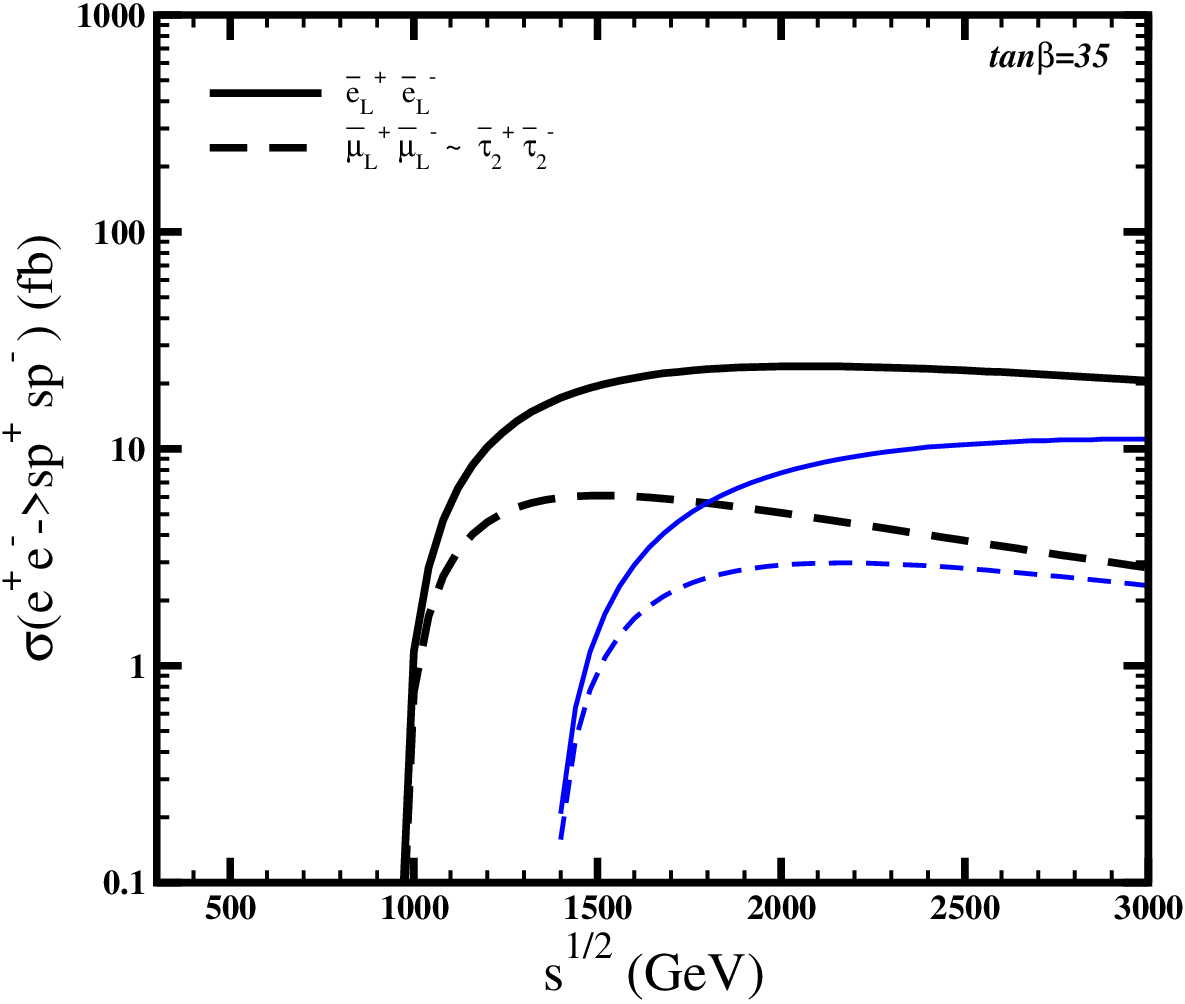} 
\includegraphics[scale=.40]{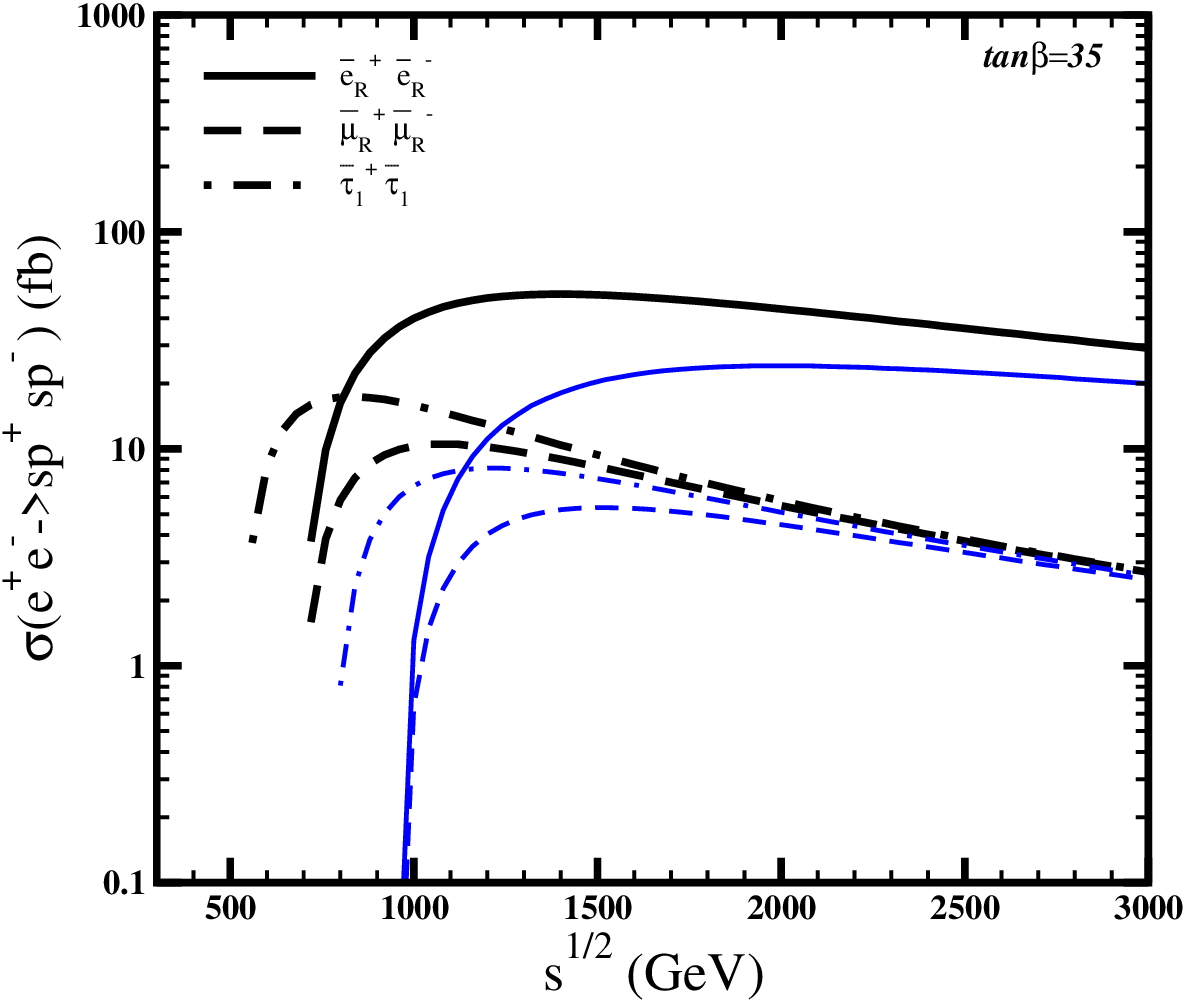} 
\includegraphics[scale=.40]{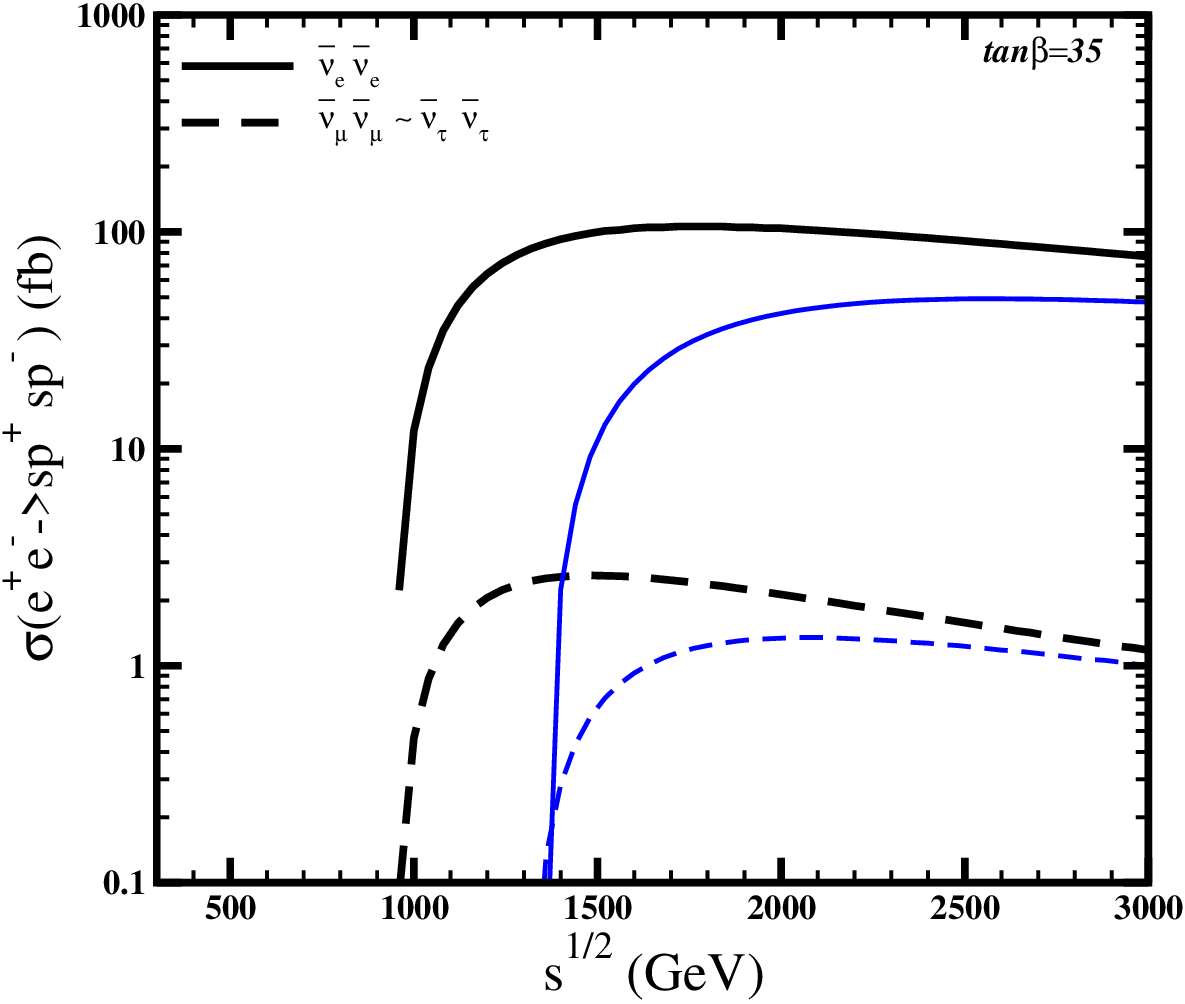} 
\includegraphics[scale=.40]{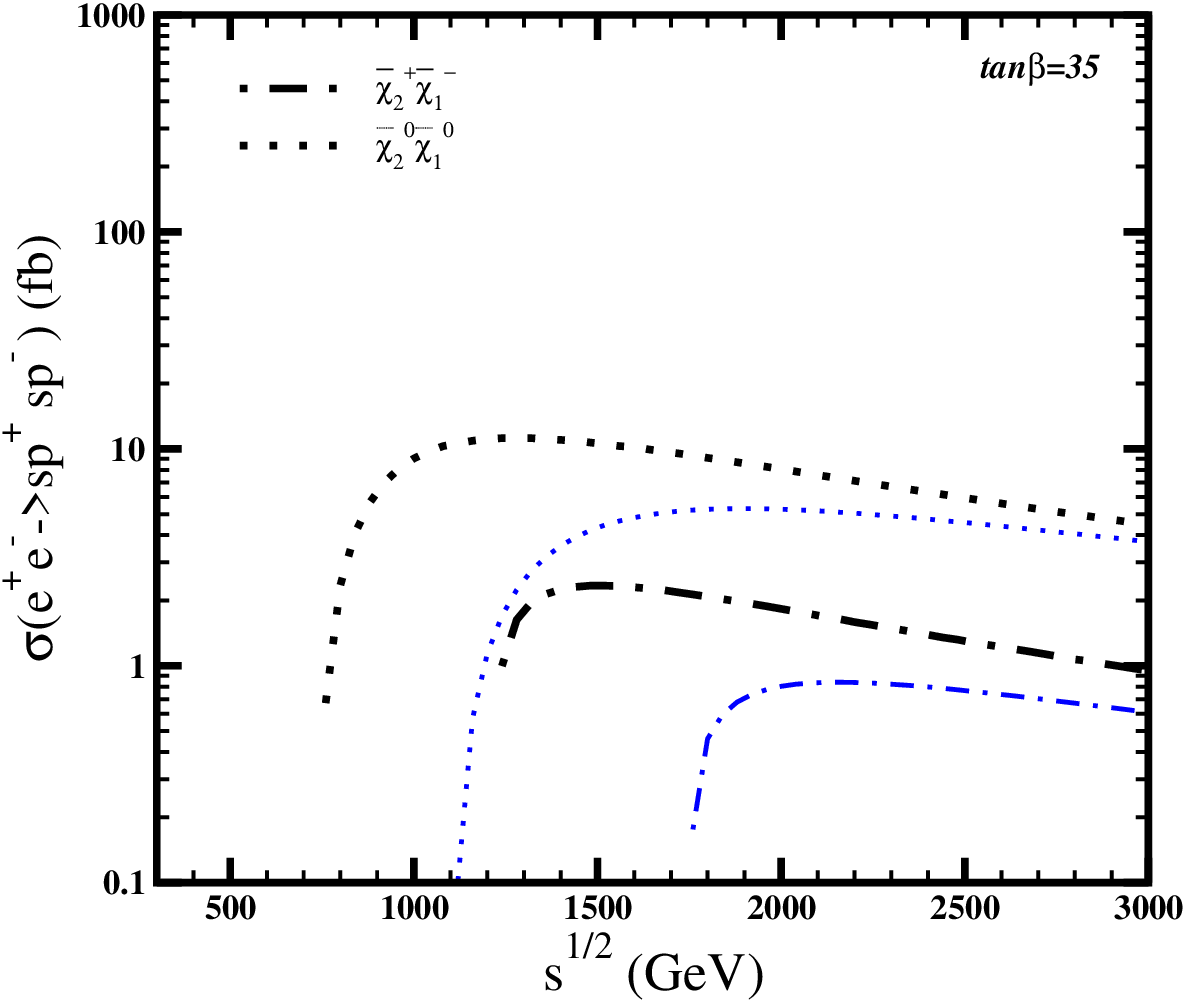} 
\end{center}
\caption{\small \it Energy dependence of the cross sections 
for sparticle production in the channels (\ref{eq:pair})
and (\ref{eq:single}). The black thick 
(blue thin) lines correspond to point a35 (b35) in (\ref{benchmarks}). 
The upper left (right) panels display left (right) charged slepton production,
and the lower left (right) panels display sneutrino (gaugino) production.
 }
\label{fig:fcsigmas35}
\end{figure}

In Fig.~\ref{fig:areaCS} we present contours of the production cross sections for sparticles that are relevant for our study of LFV effects within the CMSSM,
using 1~fb as a reference cross-section value.
We indicate in green the  constraint on the parameter space derived from the 
WMAP-favoured CMSSM parameter space  ($0.094< \Omega h^2 <
0.128$ at $3-\sigma$) using {\tt MicrOMEGAS}~\cite{MicrOMEGAs}, and in grey the regions excluded by the constraints that 
BR($b\rightarrow s \gamma $)  $> 2.5 \cdot 10^{-4}$ and $m_h>111{\rm ~GeV}$ (
we assign an uncertainty of $\sim$3~GeV to the theoretical 
calculation of $m_h$). We present contours for the production of 
$\tilde{\tau}_1^+\tilde{\tau}_1^-$ calculated using {\tt CalcHEP}~\cite{CalcHEP}: the cross sections for producing other
right sleptons would have similar contours, though at somewhat smaller values of $(M_{1/2}, m_{0})$
because they are heavier than the $\tilde{\tau}_1$. The left slepton production
cross sections are smaller, in general, and here we draw the    
$\sigma (\tilde{\nu}_{\tau}\tilde{\nu}_{\tau}) = 1$~fb contour
(the cross sections for all slepton pair-production
processes are compared in Fig.~\ref{fig:fcsigmas35}). 
The relevant channels for 
LFV in cascade decays following pair production are also displayed in Fig.~\ref{fig:areaCS}, along with 
 the line $m_{\tilde{\chi}_1^+}=m_{\tilde{\nu}_\tau}$, above which
chargino-mediated LFV processes are allowed.

We would also like to note the following, regarding comparisons with
known results: For $\tan\beta=30$ our results compare very well with the ones 
obtained using the SUSPECT code \cite{suspect}, as implemented in 
Micromegas \cite{MicrOMEGAs}. At $\tan\beta=45$ we observe very small 
differences, due to the fact that we include
all SUSY threshold corrections to $m_b$ and $m_\tau$ (from ref.\cite{GNIS1}). 
Note that the resonances on the $\chi$ annihilation channels, relevant
at large  $\tan\beta$, are dependent on  the value of $\lambda_b$, and
therefore on the  accuracy of the approximation in computing $m_b$. 
To quantify the difference, we note that our results for 
$\tan\beta=45$ are similar to the ones obtained with suspect for 
$\tan\beta=46$ and $m_b(m_b)$=4.4 GeV.

Focusing initially on the lines corresponding to  $\sqrt{s}=500{\rm ~GeV}$, we see that on-shell production of left sleptons would occur mostly in the region
excluded by $b\rightarrow s \gamma$.
At this energy, therefore, we expect 
right slepton production to be more important. 
However, production of charged slepton pairs in the cascade 
decays of neutralinos is also possible, whilst chargino decays are disfavored.
At $\sqrt{s}=1000{\rm ~GeV}$, we see a significant part of 
the parameter space with left slepton pair production (below the dashed line) and 
with both left and right slepton production (below the solid line). 
The production of pairs of charged sleptons in chargino cascade decay is not possible in most of the 
 WMAP favored area, since $m_{\tilde{\chi}_1^+}<m_{\tilde{\nu}_\tau}$.
Finally, 
at $\sqrt{s}=2000{\rm ~GeV}$ all the decay channels are present
 in the region of parameter space favoured by WMAP.

Based on this discussion, two benchmark points in the allowed regions for each of the choices
$\tan \beta = 35, 45$ have been selected for subsequent study, namely:
\bea
(a35)~~~ &\tan\beta=35& m_0=255 \rm{~GeV},\;\;\; M_{1/2}=610 {\rm ~GeV}, \nonumber  \\ 
(b35)~~~ &\tan\beta=35& m_0=345 \rm{~GeV},\;\;\; M_{1/2}=900 {\rm ~GeV}, \nonumber \\ 
(a45)~~~ &\tan\beta=45& m_0=440 \rm{~GeV},\;\;\; M_{1/2}=550 {\rm ~GeV}, 
\nonumber \\
(b45)~~~ &\tan\beta=45& m_0=730 \rm{~GeV},\;\;\; M_{1/2}=900 {\rm ~GeV} 
\label{benchmarks}
\eea
with $A_0 = 0$ in each case, resulting in the particle spectra shown in Table~1. We note that all these
points lie beyond the regions of CMSSM parameter space excluded by 2010 LHC data~\cite{LHC,MC6}.
In general, these points lead to more promising detection prospects 
at a LC at high energies:  at $\sqrt{s}=500 {\rm ~GeV}$, only production through 
right sleptons is allowed.

\begin{table}[!h]
\begin{center}
\begin{tabular}{| c  | c | c | c |}
\hline
Point & $m_0$ & $M_{1/2}$ & $\tan\beta$ \\
\hline
a35 & $255$ & $610$ & $35$ \\ 
\hline
b35 & $345$ & $900$ & $35$ \\ 
\hline
a45 & $440$ & $550$ & $45$  \\ 
\hline
b45 & $730$ & $900$ & $45$ \\ 
\hline
\end{tabular}
\begin{tabular}{| c | c | c | c | c | c | c | c | c | c | }
\hline 
$M_{\tilde{\chi}_1^+}$ & $M_{\tilde{\chi}_2^+}$ & $M_{\tilde{\chi}_1^0}$ & $M_{\tilde{\chi}_2^0}$ & $M_{\tilde{\tau}_1}$ & $M_{\tilde{\tau}_2}$ & $M_{\tilde{\ell}_R}$ & $M_{\tilde{\ell}_L}$ & $M_{\tilde{\nu}_\tau}$&  $M_{\tilde{\nu}_e}$ \\
\hline
$480$ & $741$ & $255$ & $480$ & $261$ & 
$477$ & $342$ & $478$ & $454$& $472$  \\
\hline
$720$ & $1034$ & $383$ & $720$ & $385$ & 
$670$ & $478$ & $683$ & $655$& $679$  \\
\hline
$432$ & $671$ & $229$ & $433$ & $338$ & 
$543$ & $485$ & $571$ & $519$& $566$  \\
\hline
$725$ & $1015$ & $385$ & $725$ & $590$ & 
$872$ & $801$ & $936$ & $858$& $933$  \\
\hline
\end{tabular}
\caption{\small \it Relevant sparticle masses (in ${\rm ~GeV}$) for the selected
benchmark points. The selected points all have $A_0 =0$.
}
\end{center}
\label{table1}
\end{table}

We show in Fig.~\ref{fig:fcsigmas35} the dependence of the
flavour-conserving (FC) cross sections on the $\sqrt{s}$ of the collision 
for the benchmark points
(\ref{benchmarks}) with $\tan\beta=35$. The cross sections for the benchmarks
with $\tan\beta=45$ are very similar. 
We can already get an idea of the LFV cross sections 
from  the magnitudes of these FC cross sections, by looking at
the approximate formulas 
(\ref{eq:kali1}) and (\ref{eq:kali2}), which assume 
that only 2-3 mixing is present. These 
provide an initial picture of the order of magnitude 
of LFV (up to ~40\% error), simply by convoluting the 
slepton flavour-violating decays with the FC  cross sections. 
The latter  can be extracted automatically, e.g., from {\tt CalcHEP}~\cite{CalcHEP}, allowing one to
explore quickly the relevant parameter space. If the mixing was 
indeed limited to the 2-3 sector, the accuracy would be larger,
within 10\% of the exact calculation discussed at a later stage.  

As shown below,  taking into account the full flavor mixing in the 
vertex \cite{LFVhisano,LFV-LC} leads to additional channels and an enhancement in the cross sections
$\sigma(e^+e^-\rightarrow \tilde{\ell_i}\tilde{\ell_j} )$, also for the case of 
mixing in the LL sector, despite the fact that 
the mixing with the first generation is very 
constrained by the recent MEG bound
BR($\mu\rightarrow e \gamma$)$< 2.4\cdot 10^{-12}$~\cite{Adam:2011ch}.

The cross sections $e^+ e^- \rightarrow  \tilde{\ell}_i^{-}
\tilde{\ell}_i^{+}$, $i=L,1,R,2$, which are relevant for
calculating 
$e^+ e^- \rightarrow  \tilde{\ell}_i^{-} \tilde{\ell}_j^{+} \rightarrow 
\tau^{\pm} \mu^{\mp} \tilde{\chi}^{0}_1 \tilde{\chi}^{0}_1$,
are displayed on the upper panels of
Fig.~\ref{fig:fcsigmas35}.
We see that the production rates for left and right sleptons are comparable. 

In the case of
$e^+ e^-  \rightarrow  \tilde{\nu}_i \tilde{\nu}_j^c
  \rightarrow \tau^{\pm} \mu^{\mp} \tilde{\chi}^{+}_1
  \tilde{\chi}^{-}_1$,  Fig.~\ref{fig:fcsigmas35} indicates that the production rate for sneutrinos 
can be larger than that of charged sleptons.  However,
i) even if 
$e^+ e^-  \rightarrow  \tilde{\nu}_i \tilde{\nu}_j^c$ is large, the decay 
$\tilde{\nu}_i \rightarrow l \tilde{\chi}^{-}_1$ is allowed only 
in a restricted  area of the parameter space, as we can see in 
Fig.~\ref{fig:areaCS}, and
ii) the advantage of producing the $\tilde{\chi}^\pm_1$ is not 
so important in this case, because the decay
$\tilde{\chi}^\pm_1\rightarrow \chi+q_uq_d$ is not allowed in the
range of $\tan\beta$ we favour  (we recall that in the CMSSM the NLSP is usually
a stau).

Finally, we comment on the  cascade slepton production processes of 
eq.~(\ref{eq:single}). These cases are not so promising within the CMSSM, but 
the analysis is simple, using the  flavour-conserving 
cross sections for charginos and neutralinos and multiplying 
by the LFV branching ratio.
The LFV signal arises from the decay of the $\chi_2^0$ or $\chi_2^+$, 
which are produced at lower rates than the sleptons. The LFV decay  of the
$\chi_2^0$  has been already studied in the context of the LHC  in~\cite{CEGLR2}. 
Here, in agreement with~\cite{CEGLR2},
 we find that mixing in the $RR$ sector may lead to larger signals. However,
the  resulting cross sections are smaller than those mediated by sleptons. 
Also, the channel with  $\chi_2^+$  is not so relevant,  not only 
because the production of charginos is lower than that of neutralinos 
(as can be seen in Figs.~\ref{fig:fcsigmas35}),
 but also because it is only  sensitive to $LL$ mixing in points where 
$m_{\tilde{\chi}_1^+}< m_{\tilde{\nu}_\tau}$.

Other production channels arising from $\tilde{\chi}_2^\pm\tilde{\chi}_2^\mp$ and $\tilde{\chi}_i^0\tilde{\chi}_j^0$, where $i,\;j$ denote heavier neutralinos, are also open at higher energies and 
can contribute to the LFV signals. However, at the points we consider their production is 
lower than the one we study in detail and do not give any additional insight.


\section{Opportunities for LFV observation at the LC}

LFV effects are severely constrained by decays of the form BR($ \ell_ i \rightarrow  \ell_ j \gamma$),
which limit significantly the allowed parameter space, 
with respect to both the mixing  terms $\delta_{LL,RR}$ and the values 
of the soft supersymmetry-breaking masses. These limit the range of parameters in which we can
get observable cross sections without 
violating the LFV bounds. 
In general, we may expect that 
LFV decays  are progressively suppressed for a heavier sparticle spectrum at fixed values of
$\delta_{LL,RR}$, and hence that one may get observable effects only for larger values of
$\delta_{LL,RR}$. However, since the LC gives direct access to energies that are relatively high, the
suppression induced by a heavier spectrum may be smaller than those
in rare LFV decays and $\mu - e$ conversion. It is natural, therefore, to ask the following questions.
(i)  For a heavy sparticle spectrum, how small are the $\delta_{LL,RR}$ that
are accessible before the cross sections become too small to be observable? 
(ii) For a light sparticle spectrum, what are the minimum values
of $\delta_{LL,RR}$ that can be probed? 

\begin{figure}[!t]
\begin{center}
\includegraphics[scale=.40]{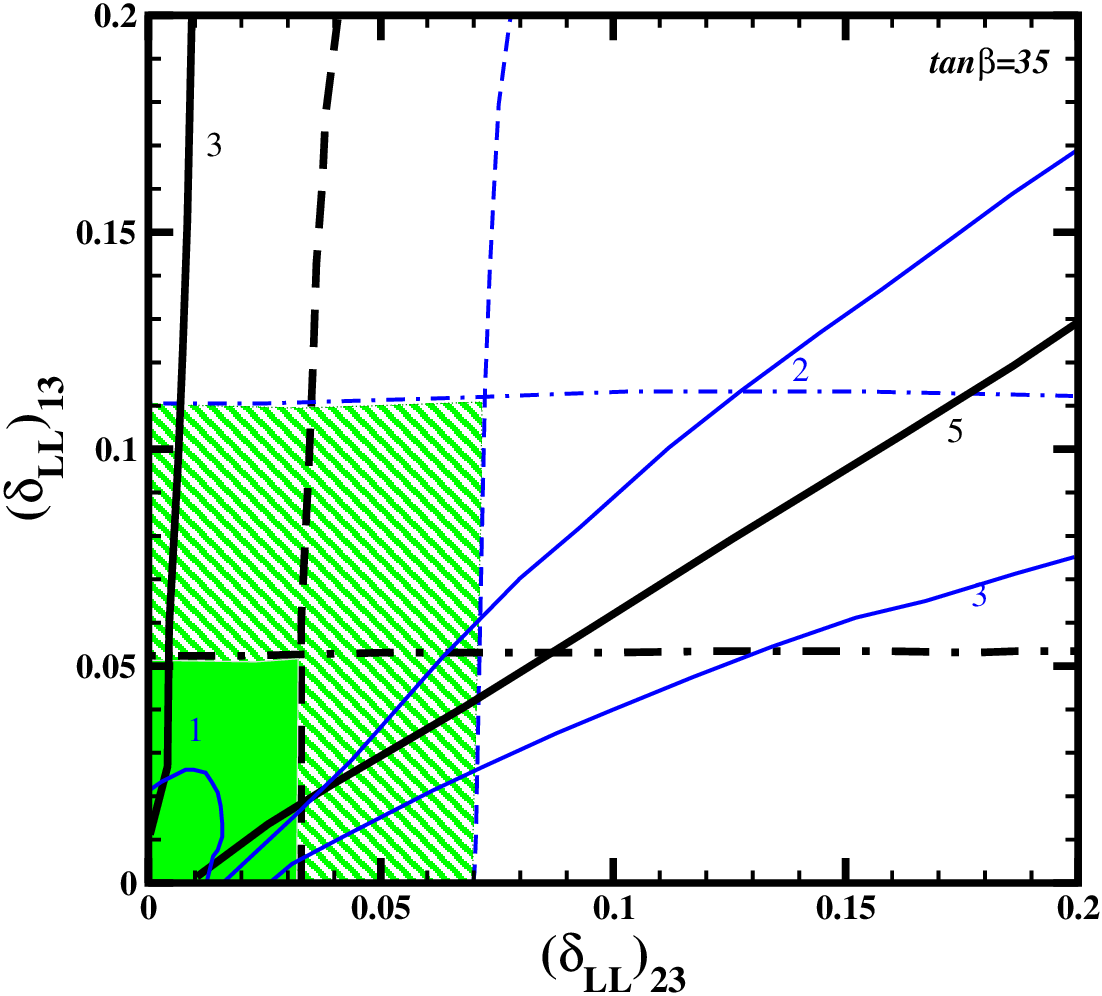} 
\includegraphics[scale=.40]{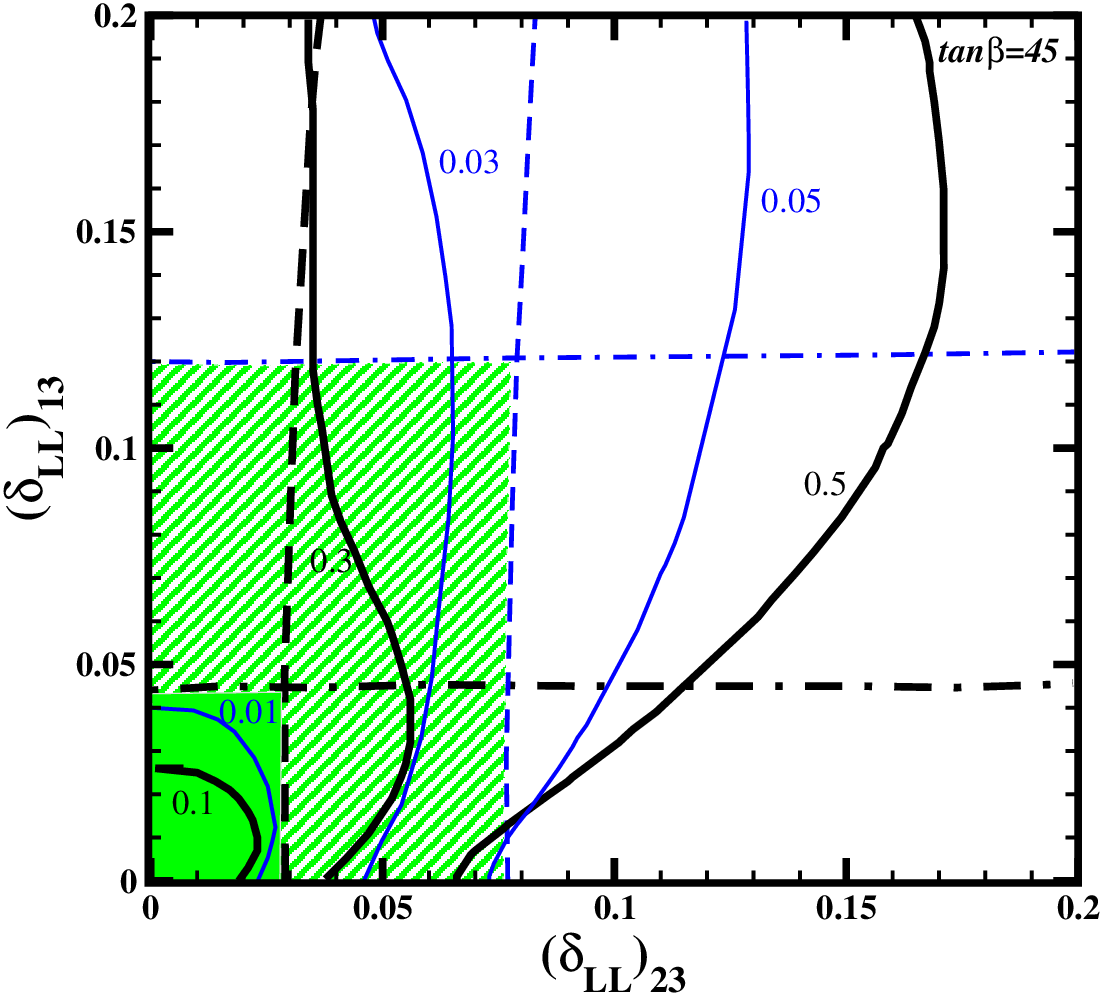} 
\includegraphics[scale=.40]{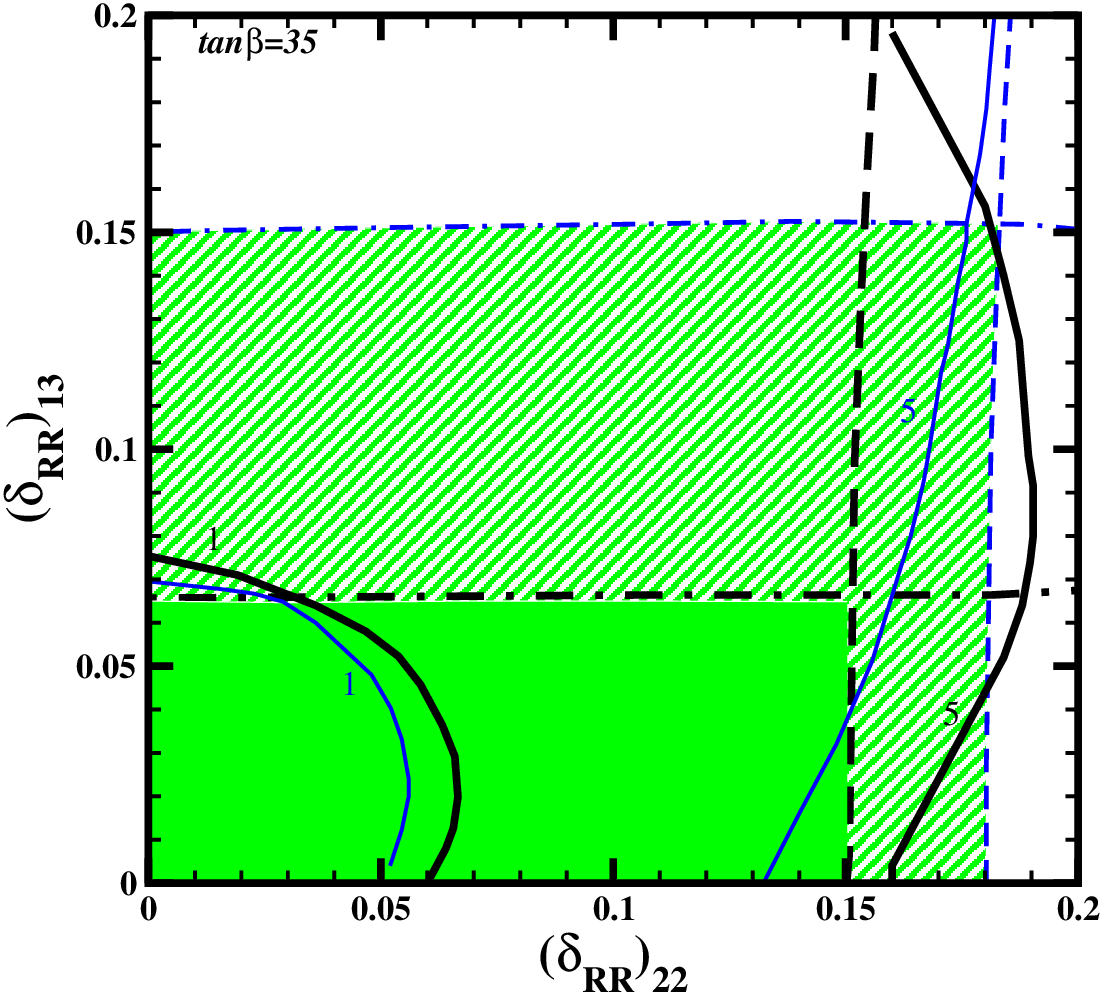} 
\includegraphics[scale=.40]{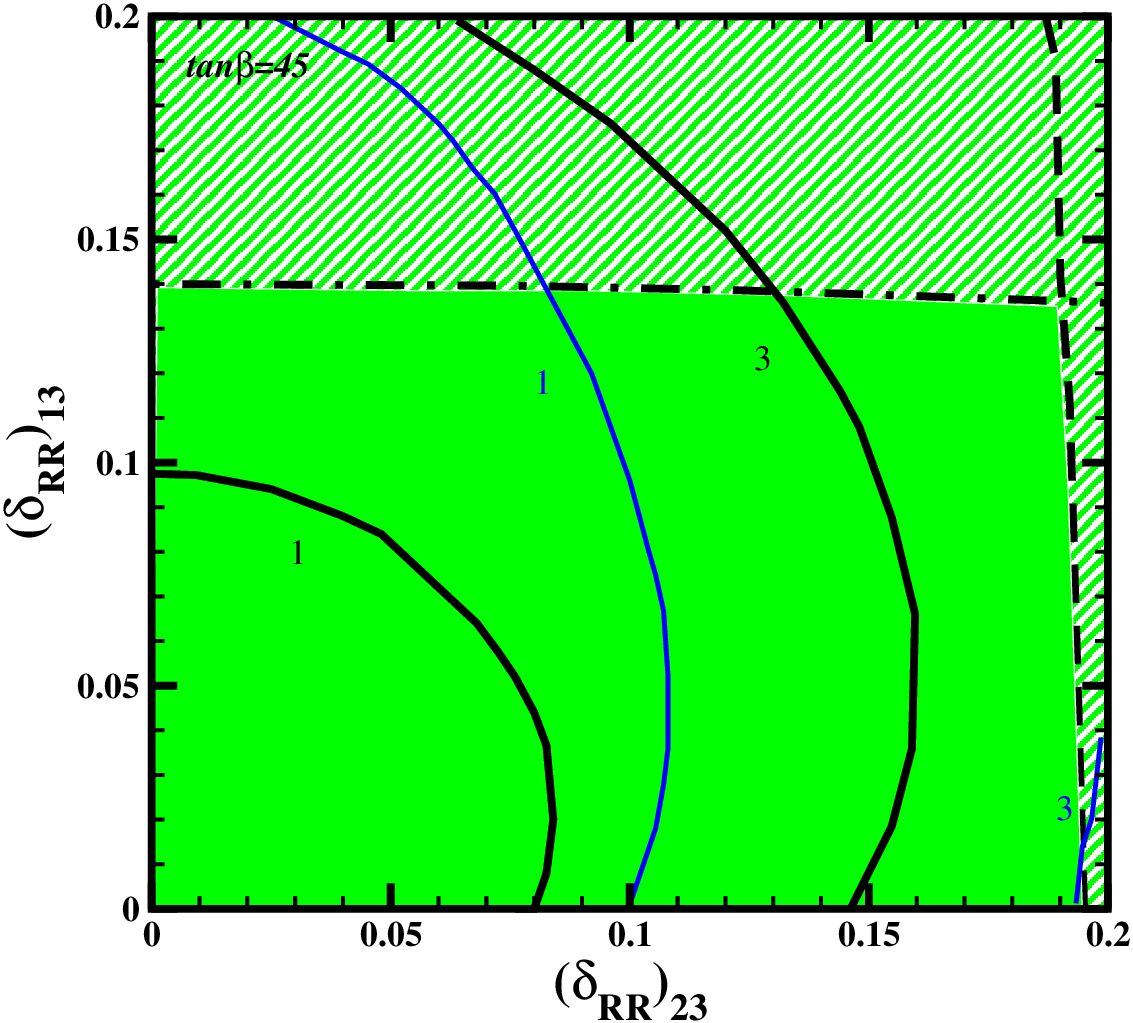}
\end{center}
\caption{\small \it Constraints on the magnitudes of the
mixing parameters and possible LFV effects for points with $\tan\beta=35$ (left panels) and 
$\tan\beta=45$ (right panels). The shaded areas are those
allowed by current limits on 
BR($\tau\rightarrow e \gamma$) (dot-dash line) and 
BR($\tau\rightarrow \mu \gamma$) (dash line) using 
$a35$, $a45$  as reference points  (thick lines bounding the solid shaded areas) 
and , $b35$, $b45$ (thin blue lines bounding  the ruled shaded areas). The 
solid lines are contours of $\sigma(e^+e^-\rightarrow \tau ^\pm\mu^\mp +2 \chi^0)$ in fb for
$\sqrt{s}=2000 {\rm ~GeV}$.
}
\label{deltas}
\end{figure}

\begin{figure*}[!t]
\begin{center}
\includegraphics[scale=.40]{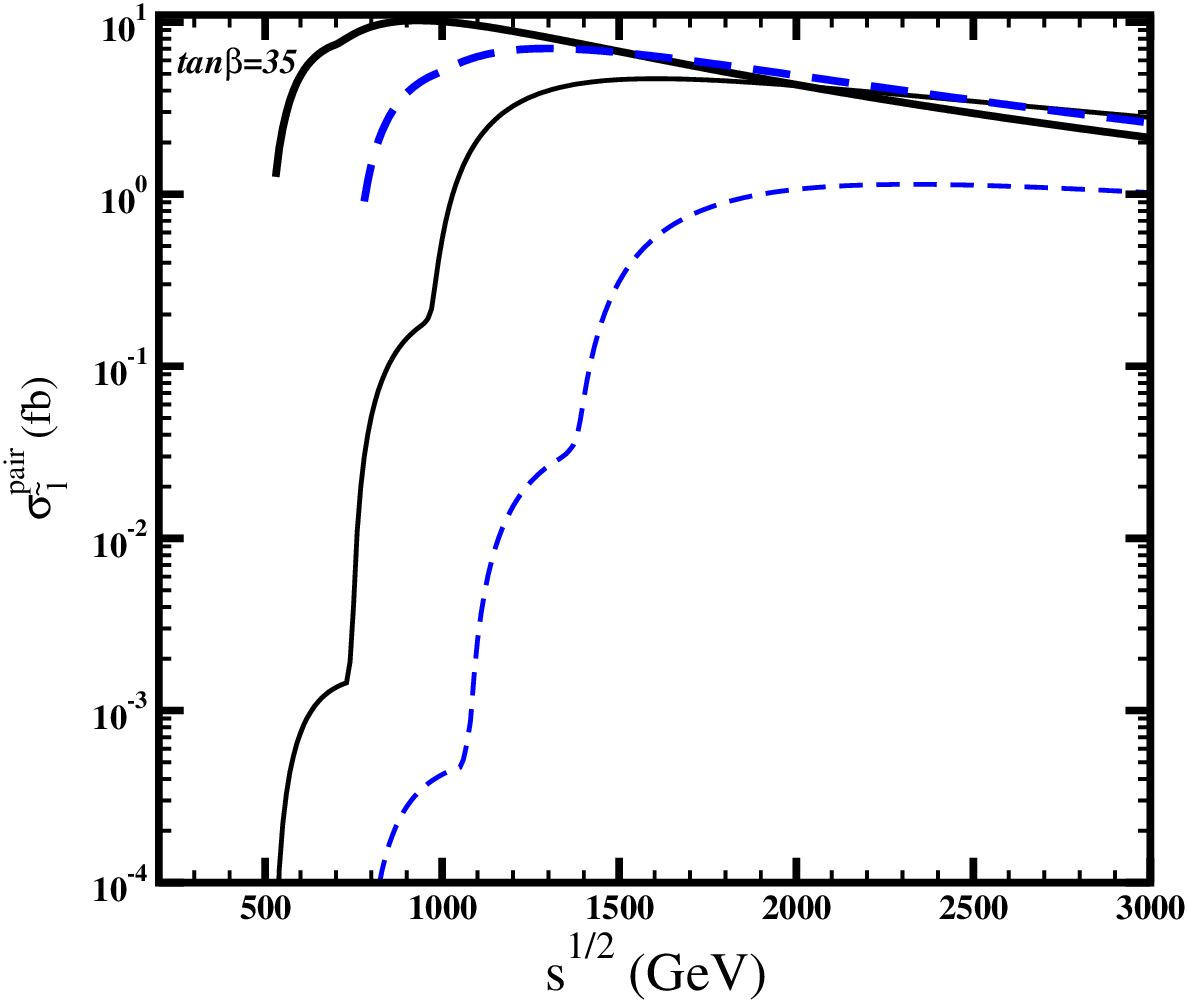} 
\includegraphics[scale=.40]{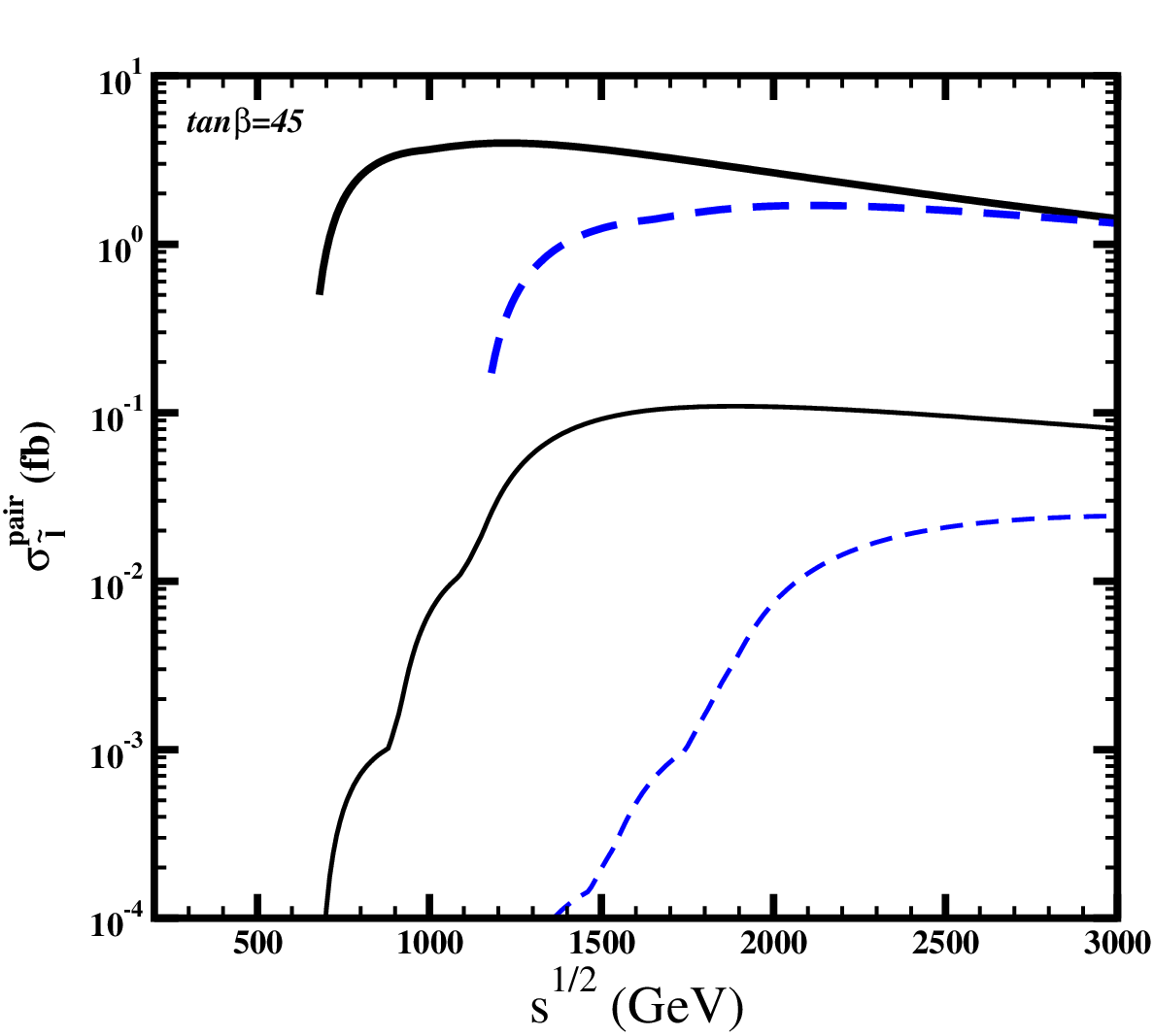} 
\end{center}
\caption{\small \it Values of $\sigma(e^+e^-\rightarrow 
\tilde{\ell}_i^{-} \tilde{\ell}_j^{+} \rightarrow \tau ^\pm\mu^\mp +2 \chi^0)$  vs $\sqrt{s}$ 
for the benchmark points with $\tan\beta=35$ (left) and $\tan\beta=45$ (right):
solid (dashed) lines  correspond to $a35$ and $a45$ ( $b35$ and $b45$). Along the 
thick lines we assume the RR mixing of (\ref{eq:choiceR}), and along the 
thin lines the LL mixing  of (\ref{eq:choiceL}).
 }
\label{fig:css_d}
\end{figure*}

These issues are addressed in Fig.~\ref{deltas}, where we display
values of the flavour-violating parameters $\delta_{RR}$ and $\delta_{LL}$, showing lines of 
constant 
$\sigma(e^+e^-\rightarrow \tau ^\pm\mu^\mp +2 \chi^0)$ for
$\sqrt{s}=2000$ ${\rm ~GeV}$~%
\footnote{We choose $\sqrt{s}=2000{\rm  ~GeV}$ in Fig.~\ref{deltas} because,
had we chosen  $\sqrt{s}=1000{\rm ~GeV}$, only the RR channels would lead to on-shell
sleptons. Even  for $\sqrt{s}=2000{\rm ~GeV}$,  we can see in Fig.~\ref{fig:css_d}
 that point $b45$ predicts very low cross sections for the LL case,
 due to the fact that the left sleptons are only marginally accessible, kinematically.}.  
 Using the bounds on 
$\tau \rightarrow \mu \gamma$ and $\tau \rightarrow \mu \gamma$ from \cite{PDG},
we display the allowed ranges in 
$((\delta_{LL})_{23}, (\delta_{LL})_{13})$ planes (upper panels), and the corresponding 
mixing parameters in the $RR$ sector (lower panels). 
In this exploration, the values of $(\delta_{LL,RR})_{12}$ were chosen to 
be proportional to those of
$(\delta_{LL,RR})_{13}$, so as to reduce the number of parameters.
We have checked that 
the dependence of our results on these parameters are small, provided they are
taken in a range such that the bound on 
BR($\mu\rightarrow e \gamma$) is preserved. In addition, we note that the ratio
 $(\delta_{LL,RR})_{13 }$($\delta_{LL,RR})_{12 }$) has a
very small influence in $\tau \rightarrow \mu \gamma$ 
($\tau \rightarrow e \gamma$).  

As we see in  Fig.~\ref{deltas}, the constraints from the LFV branching ratios result
in horizontal and vertical lines in the $\delta_{LL}$ and $\delta_{RR}$ planes, corresponding to a box of 
allowed parameters that scales with the masses in the spectrum. 
On the other hand, the lines of constant LFV cross sections depend strongly on the kinematical 
properties of the chosen point as well  as on the size of the mixing.  
In particular, we note the following:

$\bullet$ 
The shapes of the  lines are quite different for each benchmark point, as was 
to be expected from the variations 
of the cross sections between the different  benchmark points in the
WMAP favoured area. For instance,
for LL mixing, the differences between the predictions 
  for the two points at $\tan\beta=45$ (upper right panel in
 Fig.~\ref{deltas}) can be understood from the behaviours of the cross sections shown in Fig.~\ref{fig:css_d}.
 
$\bullet$ 
Except for LL mixing at $\tan\beta=45$, we obtain
LFV cross sections in the fb range
at points of the  parameter space where LFV decay branching ratios are
below the present experimental bounds. This reflects the observation that passing to
a heavier spectrum increases significantly 
the allowed size of the mixing parameters, due to the rapid decrease
of the LFV branching ratios for heavy sparticle spectra. 

$\bullet$ 
It is interesting to remark that at the LC mixing in the LL sector is
observable without 
exceeding the bounds on $ \ell_ i\rightarrow  \ell_ j \gamma$~\cite{Adam:2011ch,PDG}, 
due to the presence of additional channels as compared to
the LHC.  In the case of the LHC~\cite{CEGLR2},
mixing in  the $RR$ sector (which is less constrained but would require a
departure from the CMSSM) had turned out to be the most promising avenue
for discovering flavour violation. This is a clear distinction 
between the two colliders, and implies an advantage for the LC, in that its searches can be more
directly connected with neutrino mass and mixing parameters.

The above results motivate a particular choice of parameters for 
a more detailed discussion of LFV cross sections. We  assume 
the same values of the mixing parameters for both choices of 
$\tan\beta$, namely:
\beq
(\delta_{LL})_{13}=0.02, \; \; (\delta_{LL})_{23}=0.02,
\label{eq:choiceL}
\eeq
\beq
(\delta_{RR})_{13}=0.04, \; \; (\delta_{RR})_{23}=0.15.
\label{eq:choiceR}
\eeq 
For completeness, we also introduce a small mixing between the first 
and second generation: $(\delta_{LL/RR})_{12}=0.2\cdot
(\delta_{LL/RR})_{13}$. The LFV 
decays into $\tau-\mu$ pairs  are not heavily dependent on this parameter while, 
with this choice, the bound on BR($\mu\rightarrow e \gamma$) does not
over-constrain the parameter space.

LFV in the $\mu-e$ channel is generally 
suppressed in our scheme, unless
 $\delta_{LR}$ mixing is introduced.
In our analysis, however, we do not include such a mixing,
in order to be able to discuss separately the 
LFV arising from each chiral sector, thereby avoiding the additional model dependence
associated with choosing $\delta_{LR}\neq 0$.
In the case of large $RR$ mixing, non-trivial 
interference  of the $RR$ sector with the $LL$ sector 
would arise through $LR$ 
mixing. The $\mu\rightarrow e \gamma$ bound,
which could almost be ignored in selecting 
$RR$ and $LL$ mixing, would now play a very important role,
forcing us to either a very small part of the parameter space
or to very small values of $(\delta_{LR})$.
To reach the bound on radiative decays with $LR$ contributions alone we would 
need: 
$(\delta_{LR})_{12} \sim 10^{-3}$, $(\delta_{LR})_{13}\sim 10^{-2}$~\cite{playground}.

We also followed the conservative approach of keeping
$A_0 =0$, avoiding the introduction of other new LFV parameters.  
The main consequence for our analysis of allowing $A_0 \neq 0$ would be to allow the chiral composition of the staus to vary, since the 
first and second generations have small Yukawa couplings. 
However, additional LFV could be introduced by
assuming a non-trivial mixing matrix for the
$A$ terms. In this case, chiral
mixing would occur for all  generations,
restricting among other effects the size of the RR mixing.
The decay $\mu \rightarrow e \gamma$ is the most sensitive to LFV  $A$ terms, 
and  establishing a correlation between $\mu$ and $\tau$ decays 
without having a specific model
would be non-trivial. Nevertheless, since the bound 
on BR($\mu \rightarrow e \gamma$) imposes a very severe restriction on $(M_{LR})_{12}$, even assumptions such as  $(M_{LR})_{23}\sim (M_{LR})_{13}\sim 100 (M_{LR})_{12}$ 
would not alter our main conclusions.

\section{Study of LFV at the LC}

The next step is to understand the dependence of the results on
the available $\sqrt{s}$. Clearly, at the LC, higher values of  $\sqrt{s}$ provide 
accessible cross sections for heavier sparticle spectra,
a potential advantage of the LC over the
LHC, where it is relatively difficult to reach large values of the scalar masses. 
The variation of  $\sigma(e^+e^-\rightarrow \tau ^\pm\mu^\mp +2 \chi^0)$ with 
$\sqrt{s}$ is displayed in Fig.~\ref{fig:css_d}, where it is shown how, as $\sqrt{s}=2 P_{cm}$ increases, it 
becomes possible to produce on-shell sleptons with larger 
masses. We should underline once more 
that  the effects due to the kinematics are as relevant as those from the LFV mixing parameters. 
The shapes of the lines can be understood as follows.

$ \bullet $
The  cross sections for slepton pair production show interesting dependence on $\sqrt{s}$, 
since additional channels arise from slepton mixing.  We see in Fig.~\ref{fig:css_d} steps in the left 
slepton production cross sections, 
which appear  as the sleptons of the first two generations become kinematically
accessible. The steps corresponding to lighter generations are less pronounced in the case of $RR$ mixing
than in the $LL$ case.

\begin{figure}[!t]
\begin{center}
\includegraphics[scale=.40]{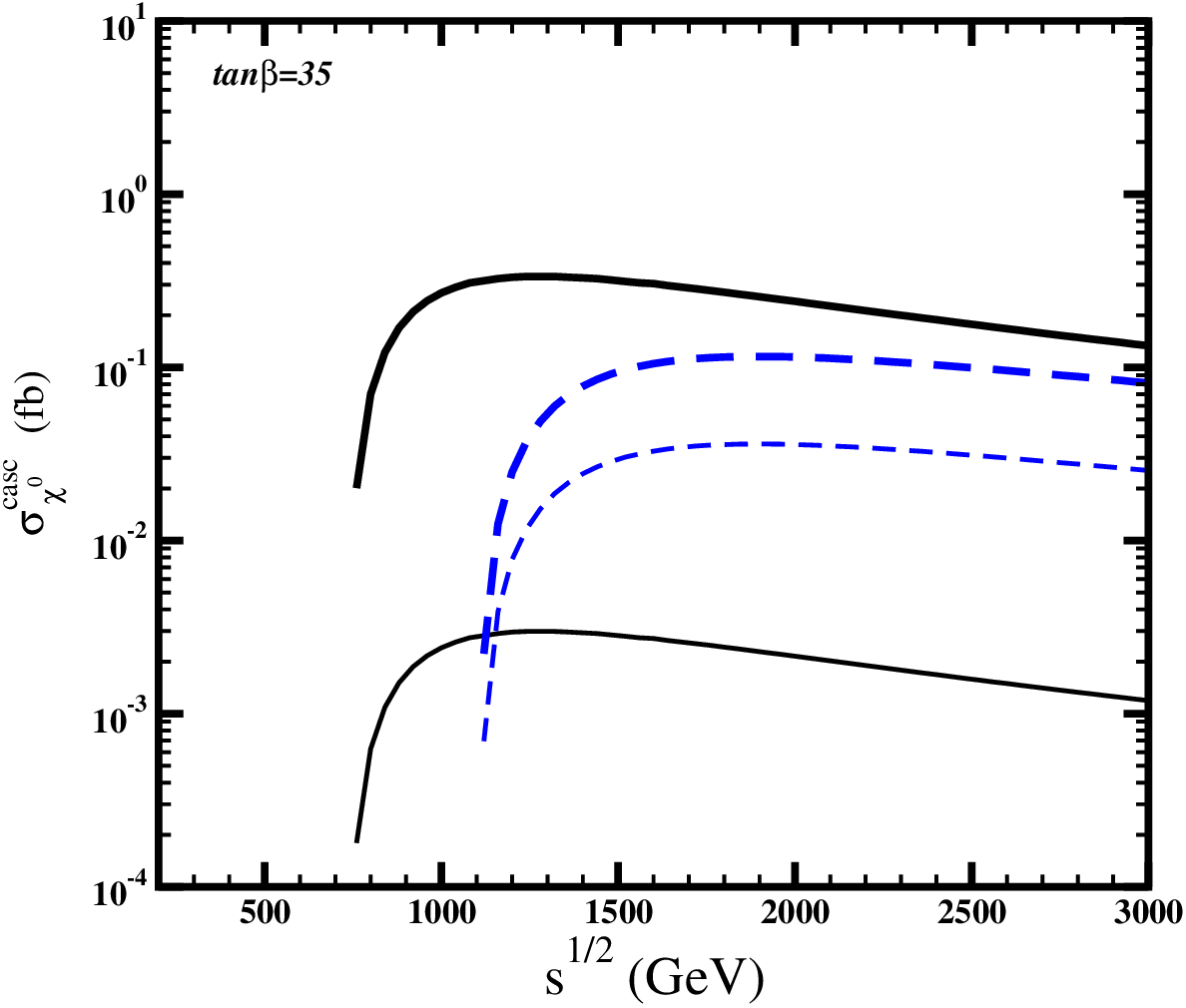} 
\includegraphics[scale=.40]{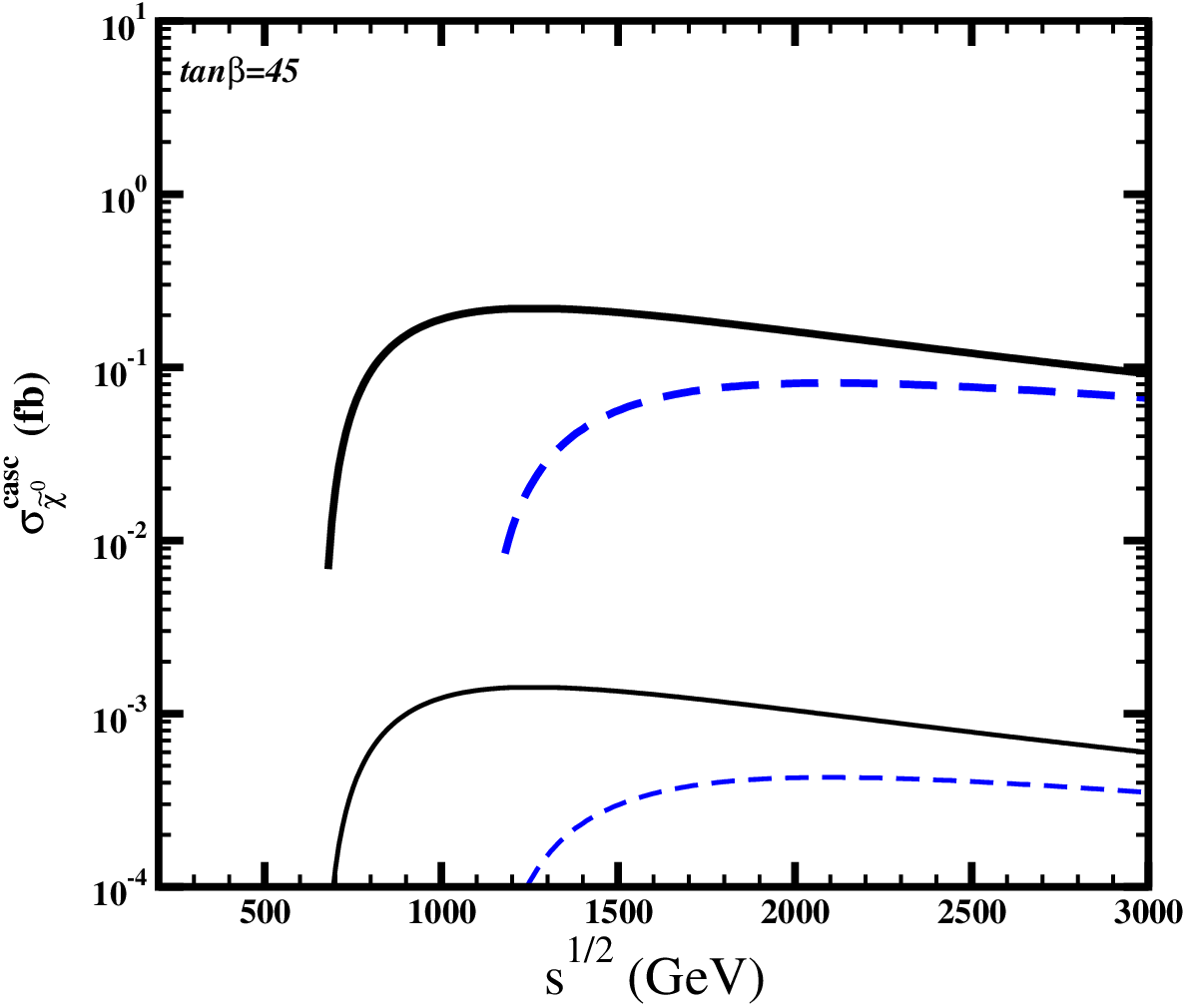} 
\end{center}
\caption{\small \it Same as in Fig.~\ref{fig:css_d}, but for cascade slepton production in the channel 
 $\sigma(e^+e^-\rightarrow \chi^0_1\chi^0_2\rightarrow \tau ^\pm\mu^\mp +2 \chi^0)$.
 }
\label{fig:css_s}
\end{figure}

\begin{figure}[t]
\begin{center}
\includegraphics[scale=.40]{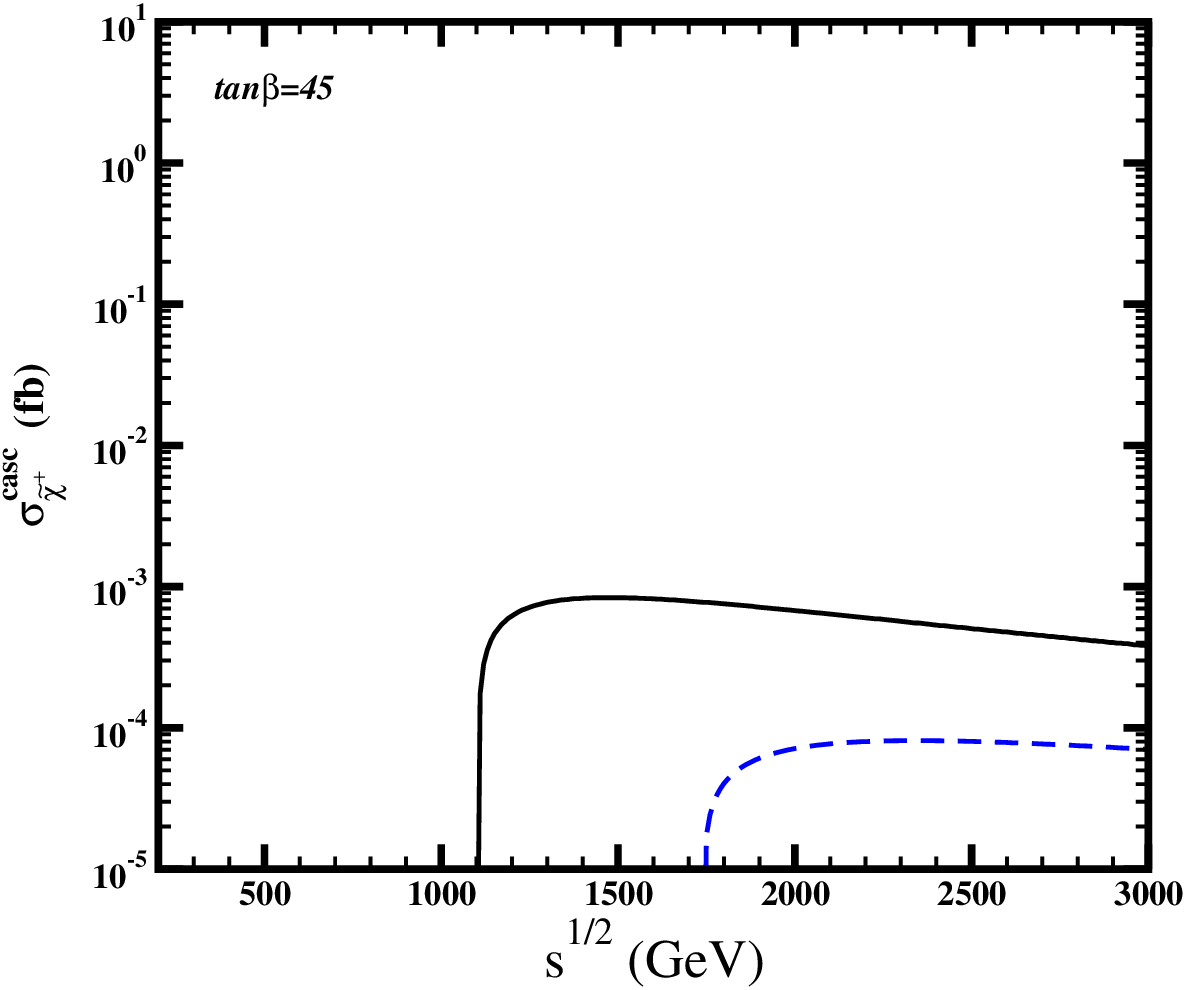}  
\includegraphics[scale=.40]{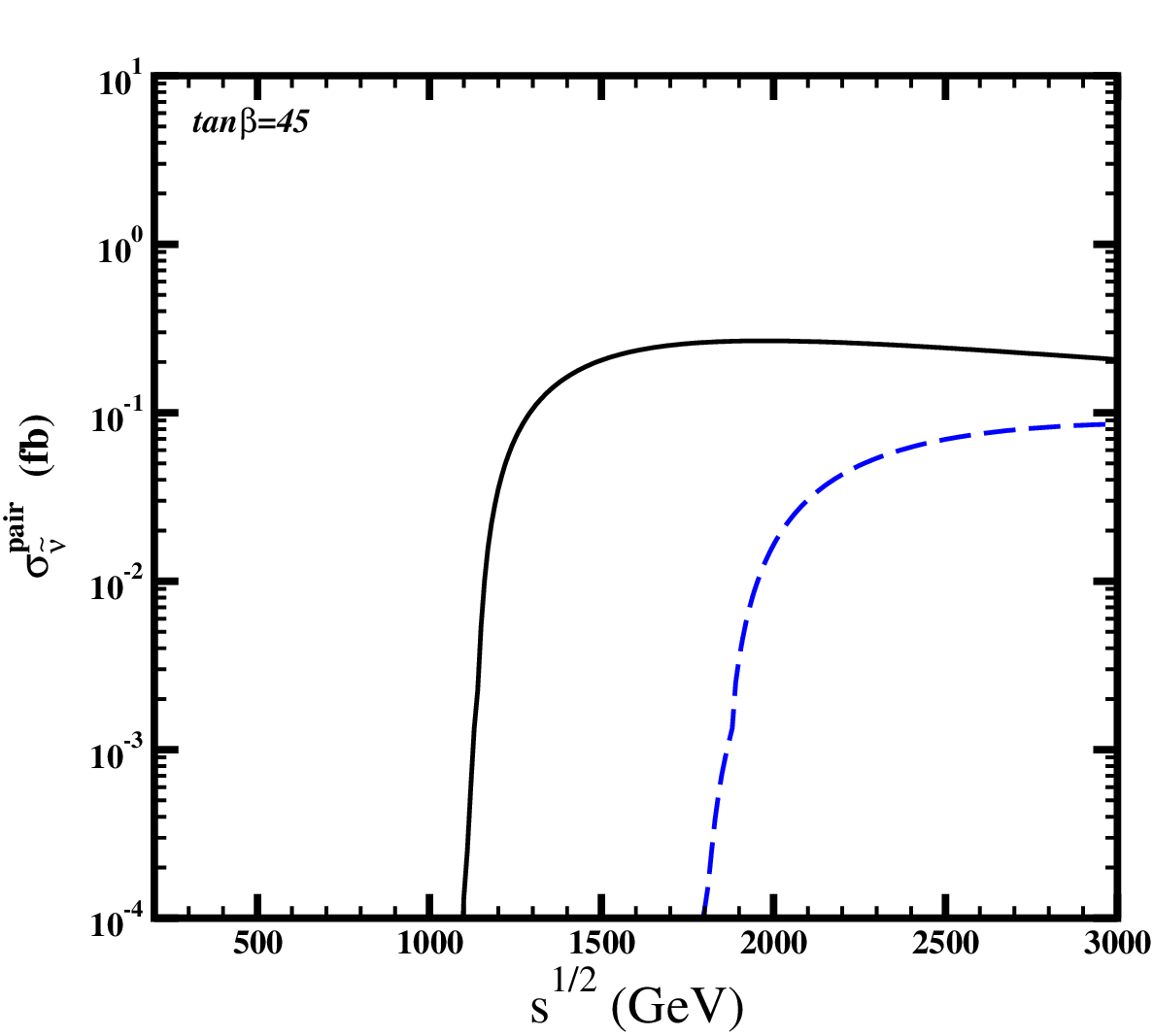}  
\end{center}
\caption{\small \it Left panel: Same as in Fig.~\ref{fig:css_d}, but for cascade slepton production in the channel 
 $\sigma(e^+e^-\rightarrow \chi^\pm_1\chi^\mp_2\rightarrow \tau
 ^\pm\mu^\mp +2 \chi^0)$. Right panel:
Same as in Fig.~\ref{fig:css_d}, but for  pair production 
with sneutrino mediation 
$\sigma(e^+e^-\rightarrow \tilde{\nu}_i\tilde{\nu}_j^c\rightarrow \tau ^\pm\mu^\mp + \tilde{\chi}_1^+ \tilde{\chi}_1^- )$.
 For $\tan\beta=35$ these channels are not allowed because $m_{\tilde{\nu}}
< m_{\tilde{\chi}^+}$.
 }
\label{fig:css_s+}
\end{figure}

$\bullet $
The overall cross sections for slepton production in cascade decays are independent 
of the LFV parameters, and LFV arises in the decays of the gauginos. 
Figs.~\ref{fig:css_s} and \ref{fig:css_s+} display the cross sections 
for the pair production of  $\tilde{\chi}_2^0 \tilde{\chi}_1^0$ or
$\tilde{\chi}_2^+ \tilde{\chi}_1^-$ 
multiplied by the LFV branching ratios for the decays of $\tilde{\chi}_2^0$ or
$\tilde{\chi}_2^+$, computed as in \cite{CEGLR2}. This explains the
 simple shapes in this case.

$\bullet$
Distinguishing  between pair and cascade slepton production
is quite crucial, particularly since, as already mentioned, it is a fundamental
difference between the LC and the LHC.
Comparing the pair and cascade slepton production cross sections in  Figs.~\ref{fig:css_d}
and  \ref{fig:css_s} (where we chose similar scales to facilitate comparisons)
we see the following: in slepton pair production, both  $LL$ and $RR$ mixing effects are of the same order of magnitude; on the contrary, in the case of 
cascade slepton production, the channel 
with $LL$ mixing has a cross section at least one order of magnitude less than that with
$RR$ mixing, offering a distinction between the two mixing scenarios.

$\bullet$
Channels with charginos in the final states could be promising
when $m_{\tilde{\chi}_1^+}<m_{\tilde{\tau}_1}$, i.e.,
when the ${\tilde{\chi}_1^+}$ is the NLSP. However, as already remarked, this is not the  
case of the CMSSM. In general, the dominant chargino 
decay is $\tilde{\chi}_1^+\rightarrow \tilde{l} \nu$, thus we do not
obtain additional information beyond that from simple slepton production and decay.
This is true also for channels with the heavier neutralinos.

$\bullet$
Cascade slepton production mediated by charginos is possible only with mixing in the 
$LL$ sector. Fig.~\ref{fig:css_s+} (left panel) indicates that this channel could
be  relevant at $\sqrt{s}$ larger than in the 
neutralino mediated channels, and the cross sections would be of
comparable magnitude in the $LL$ mixing case.
On-shell production requires 
$m_{\tilde{\nu}}>m_{\tilde{\chi}^+_1}$, which is not
the case for the sample points with $\tan\beta=35$, 
as they lie in the coannihilation region of the 
WMAP allowed area. However, for $\tan\beta=45$ the WMAP region includes points with 
resonant annihilation of neutralinos where the condition $m_{\tilde{\nu}}>m_{\tilde{\chi}^+_1}$ is satisfied. 

$\bullet$
The expectations for sneutrino pair production and subsequent chargino production are
shown in the right panel of
Fig.~\ref{fig:css_s+}. Relaxing the assumptions of the CMSSM can change the kinematics of 
$e^+e^-\rightarrow \tilde{\chi}^\pm_1\tilde{\chi}^\mp_2$, as
well as the mass ratio of $\tilde{\chi}^+_2$ versus $\tilde{\nu}$ that determines the decay 
$\tilde{\chi}^\pm_2\rightarrow \tilde{\chi}^\mp_1 \tau^\pm\mu^\mp$. The branching ratios of the  $\tilde{\chi}^\pm_1$ decays to $\nu_\tau \tilde{\tau}_1$ are 80\% and 65\% at points $a45$ and $b45$ respectively, with a subsequent $\sim 100$~\% decay of the stau into a $\tau$ and a $\chi_1^0$.

\begin{figure}[!t]
\begin{center}
\includegraphics[scale=.40]{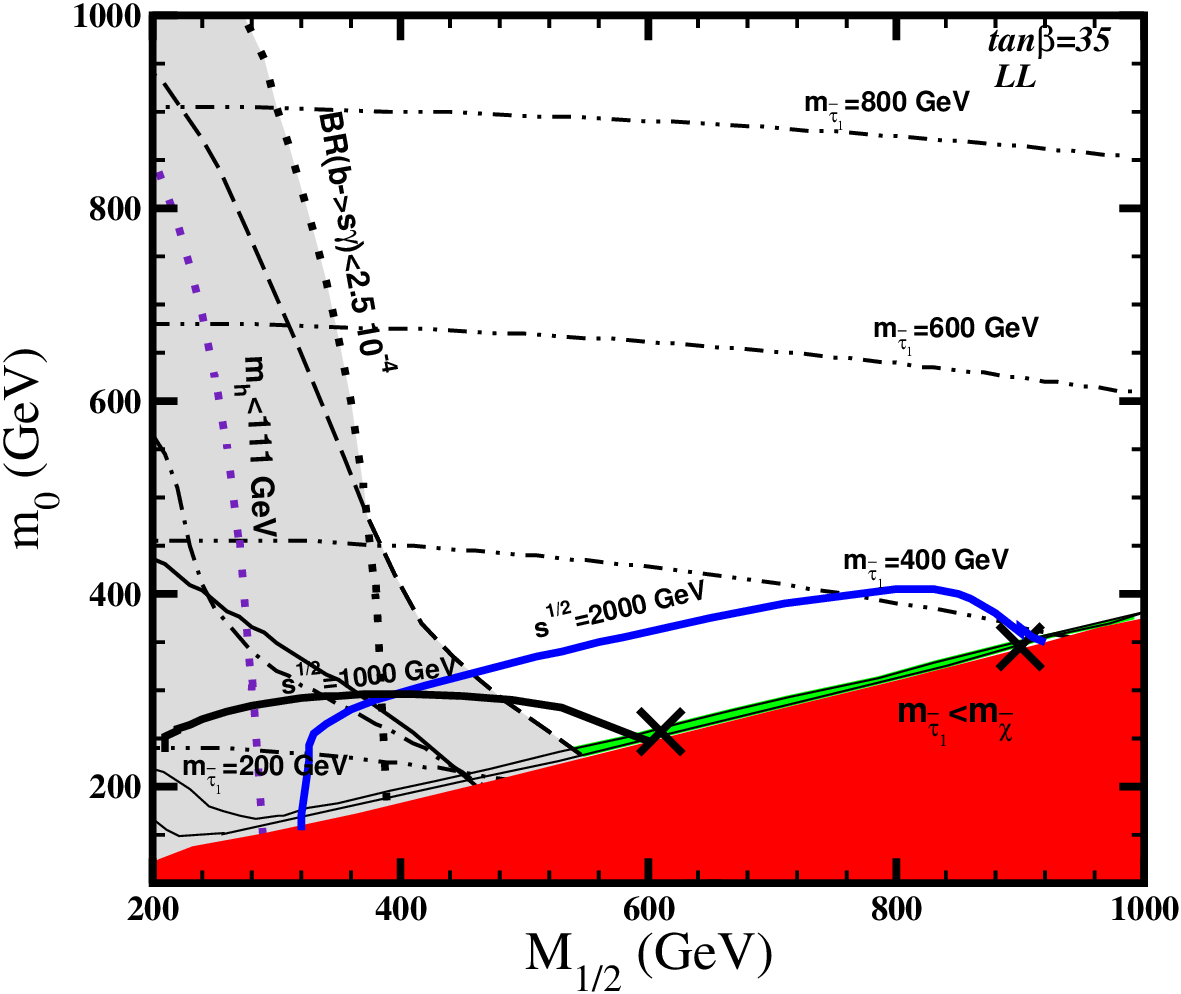}
\includegraphics[scale=.40]{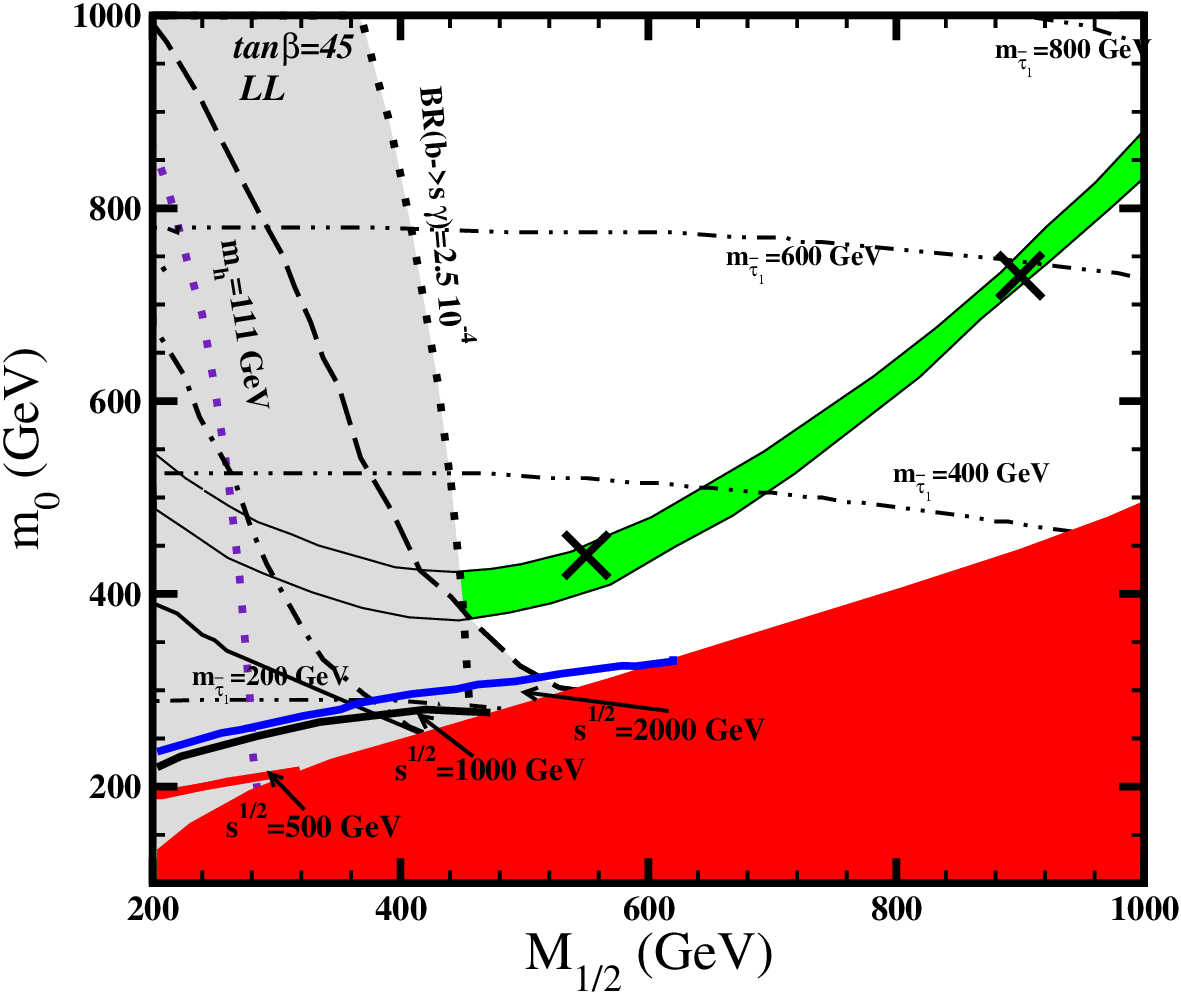}
\includegraphics[scale=.40]{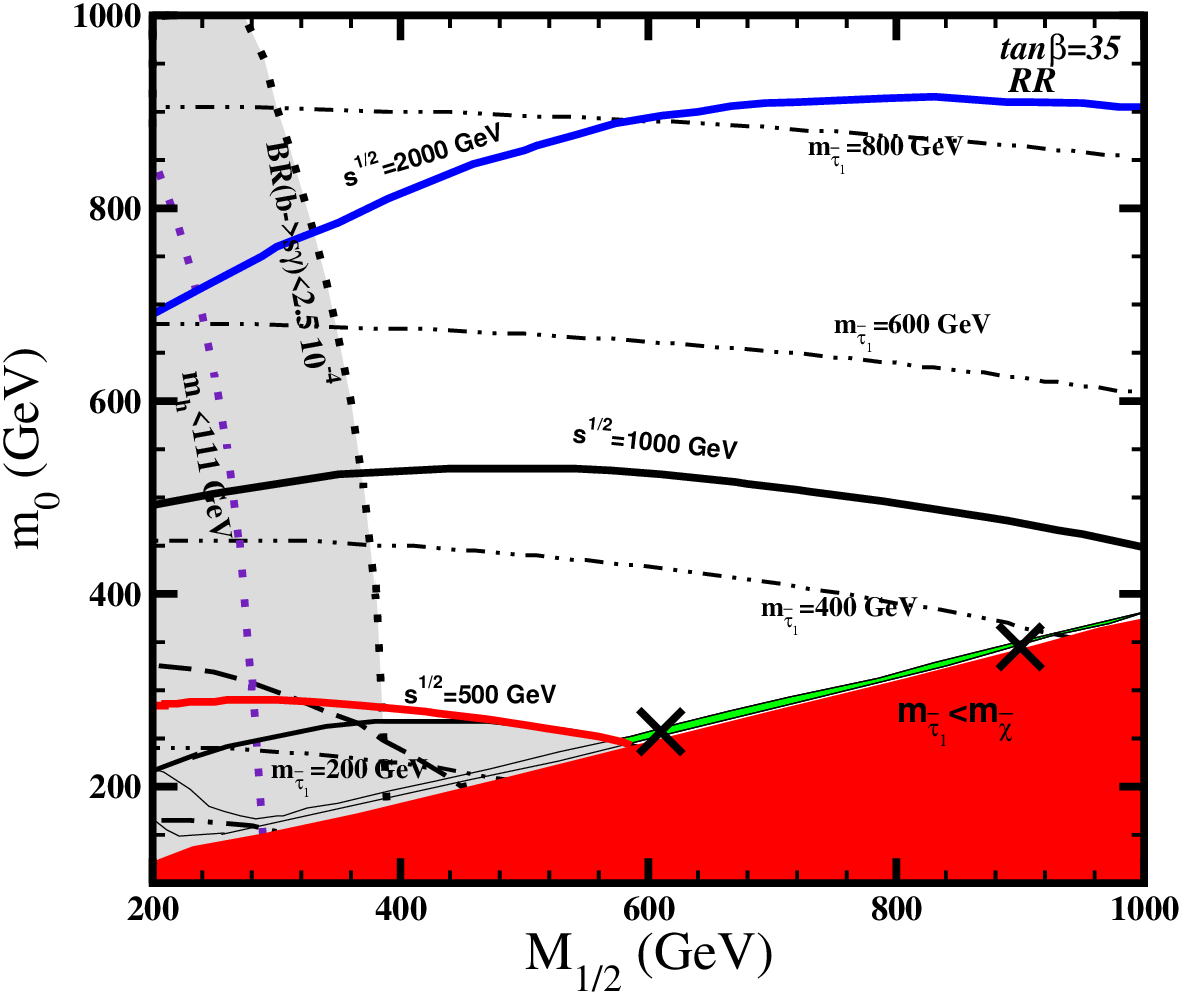}
\includegraphics[scale=.40]{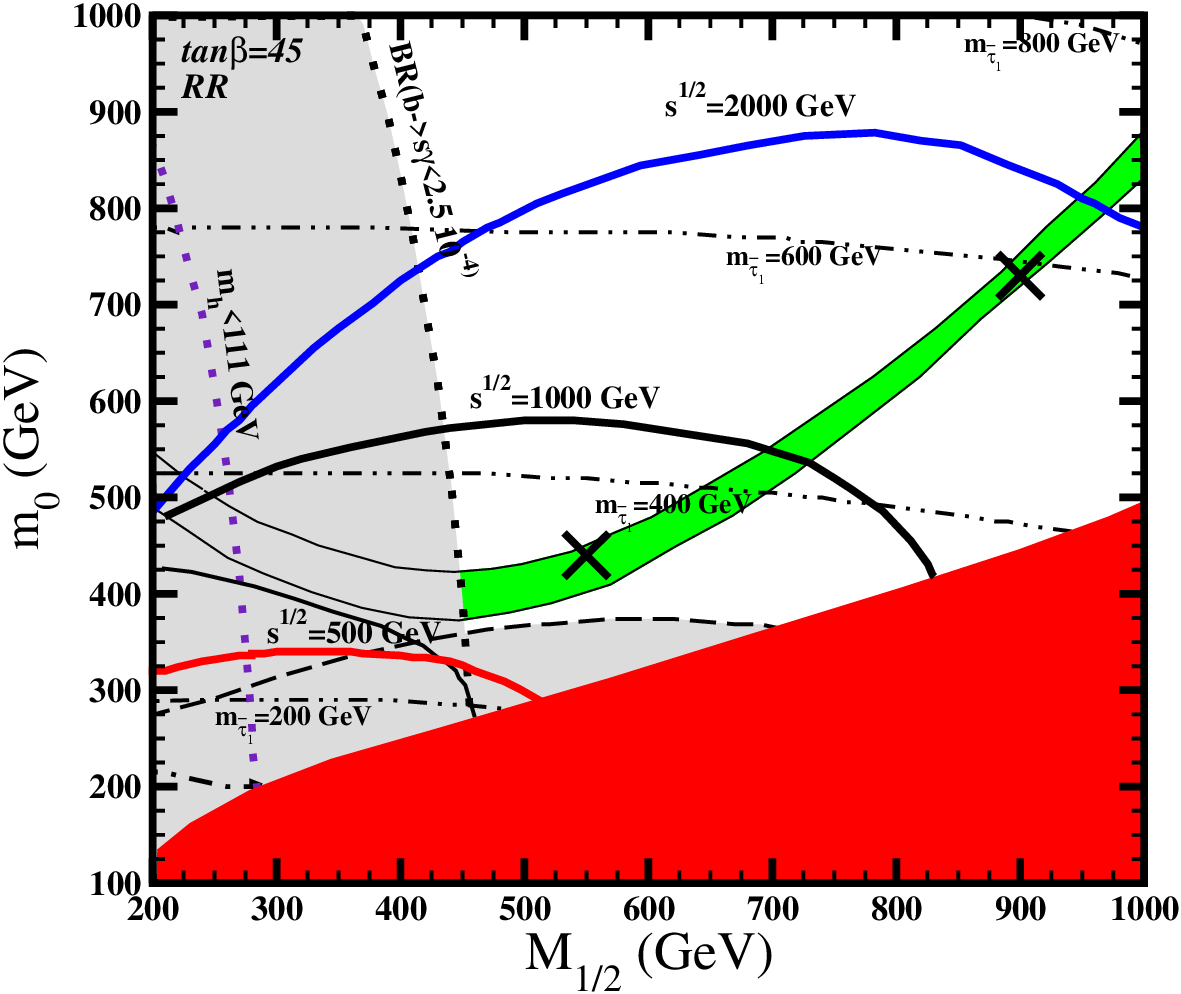}
\end{center}
\caption{\small \it The solid red (black) (blue) lines are contours
where $\sigma(e^+e^-\rightarrow \tau ^\pm\mu^\mp +2 \chi^0)= 1$~fb at 
$\sqrt{s}=500, 1000, 2000{\rm ~GeV}$, assuming the left mixing parameters 
(\ref{eq:choiceL}) in the upper panels and the  right mixing (\ref{eq:choiceR}) in the lower panels. 
The left panels are for for $\tan\beta=35$ and the right panels for $\tan\beta=45$,
assuming $A_0=0$ in both cases. Each panel also shows the 
areas excluded by current bounds on 
BR($\mu \rightarrow e \gamma$) (thin solid line),
 BR($\tau \rightarrow e \gamma$) (thin dash line), 
 BR($\tau \rightarrow \mu \gamma$) (thin dot-dash line). 
The areas excluded by BR($b \rightarrow s \gamma$) and the LEP Higgs search
are also displayed, and the green area denotes the WMAP favored region.
We see that there are ample opportunities for LFV discovery and measurement at the LC, and
the benchmark points chosen for further studies are indicated by crosses.
} 
\label{fig:areas}
\end{figure}

Cross sections for $e^+e^-\rightarrow \tau ^\pm \mu^\mp +2 \chi^0$ above 
1~fb can be reached in the areas of parameters shown in Fig.~\ref{fig:areas} for energies below those indicated.
The large areas between the thick solid red, black and blue lines, on the one hand, 
and the shaded regions that are excluded by the indicated present experimental constraints, on the other hand,
demonstrate that there are ample opportunities for LFV discovery and measurement at the LC.
These opportunities are exemplified by the benchmark points marked by crosses in Fig.~\ref{fig:areas}.

\begin{figure}[!t]
\begin{center}
\includegraphics[scale=.40]{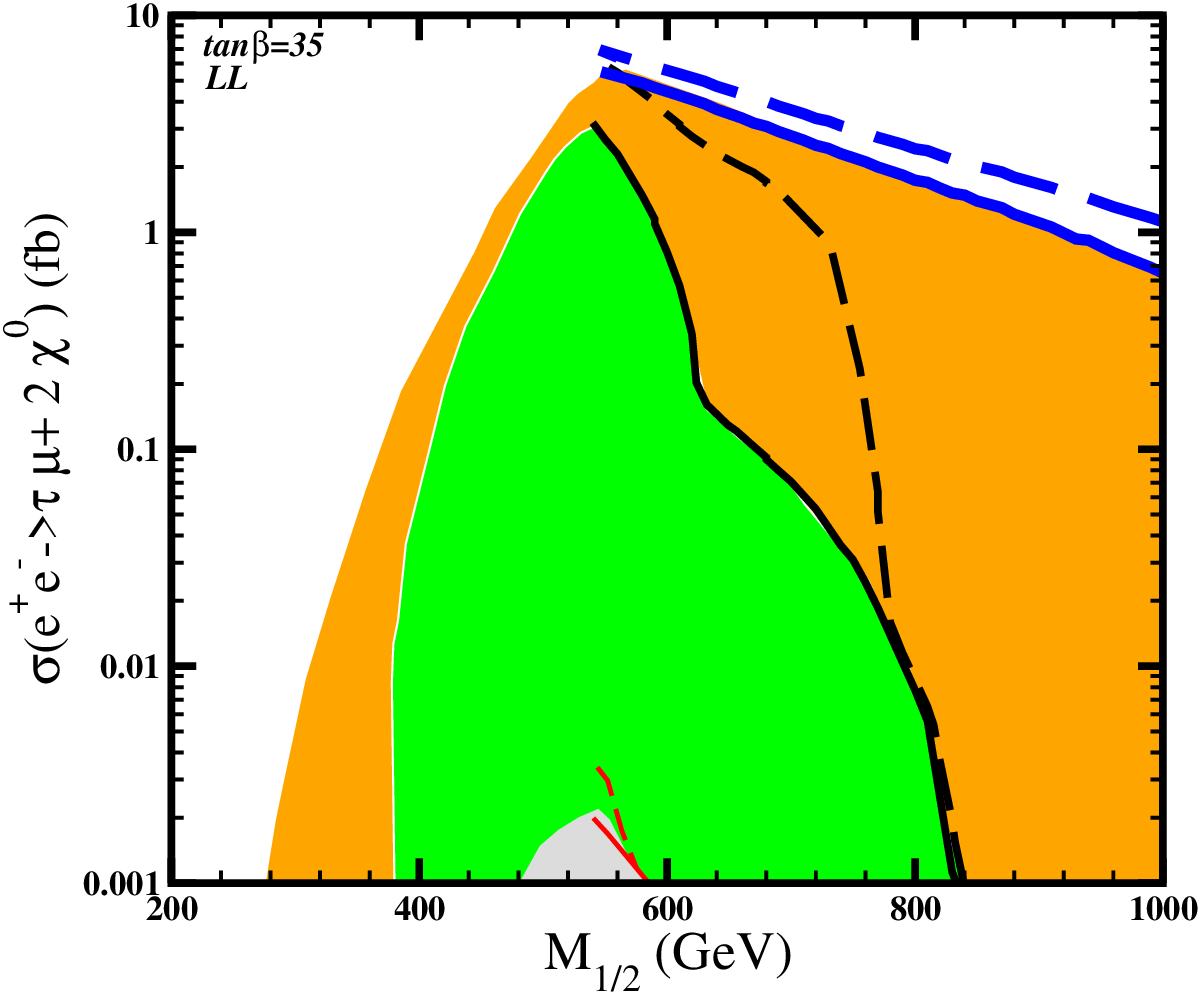}
\includegraphics[scale=.40]{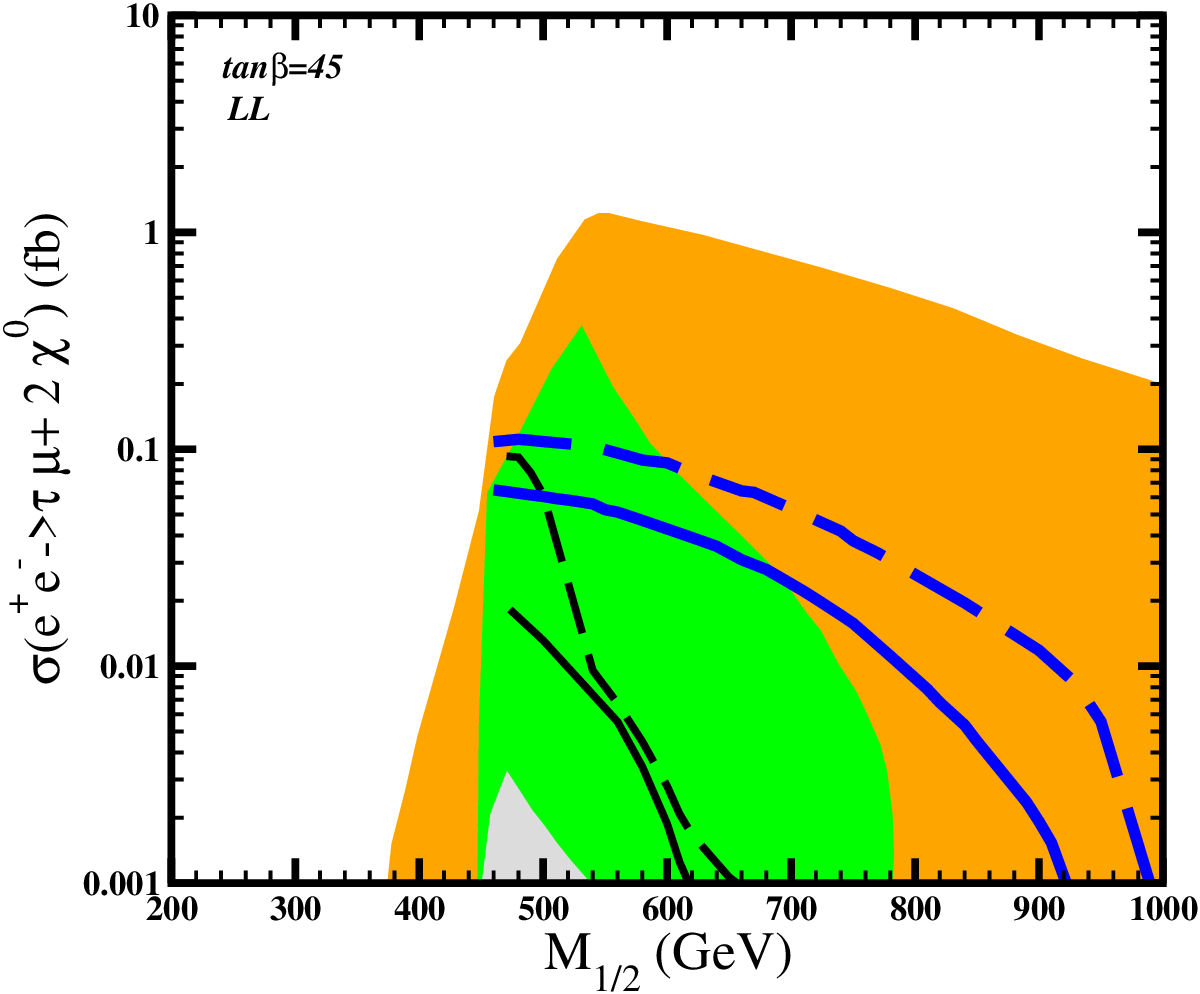}
\includegraphics[scale=.40]{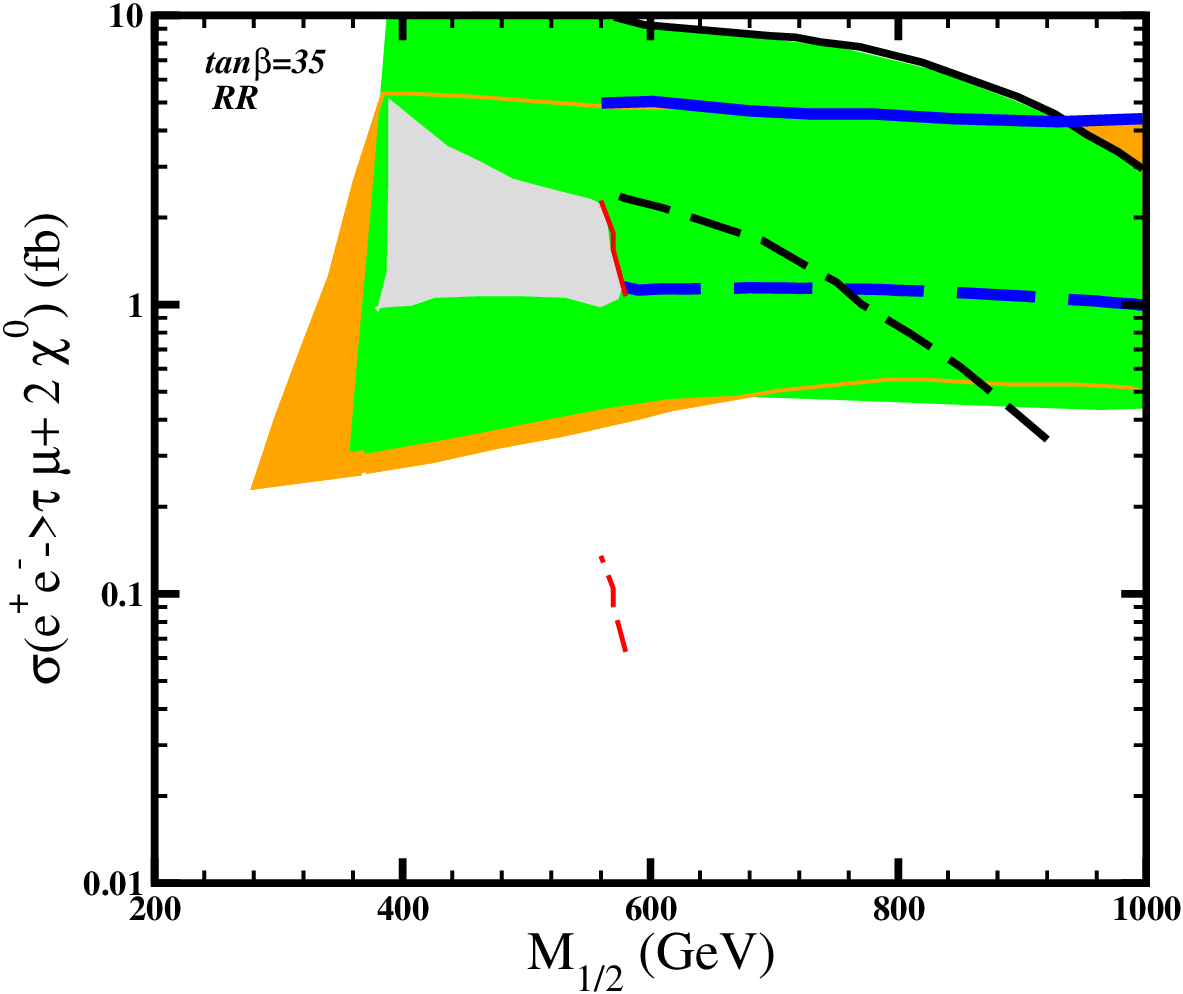} 
\includegraphics[scale=.40]{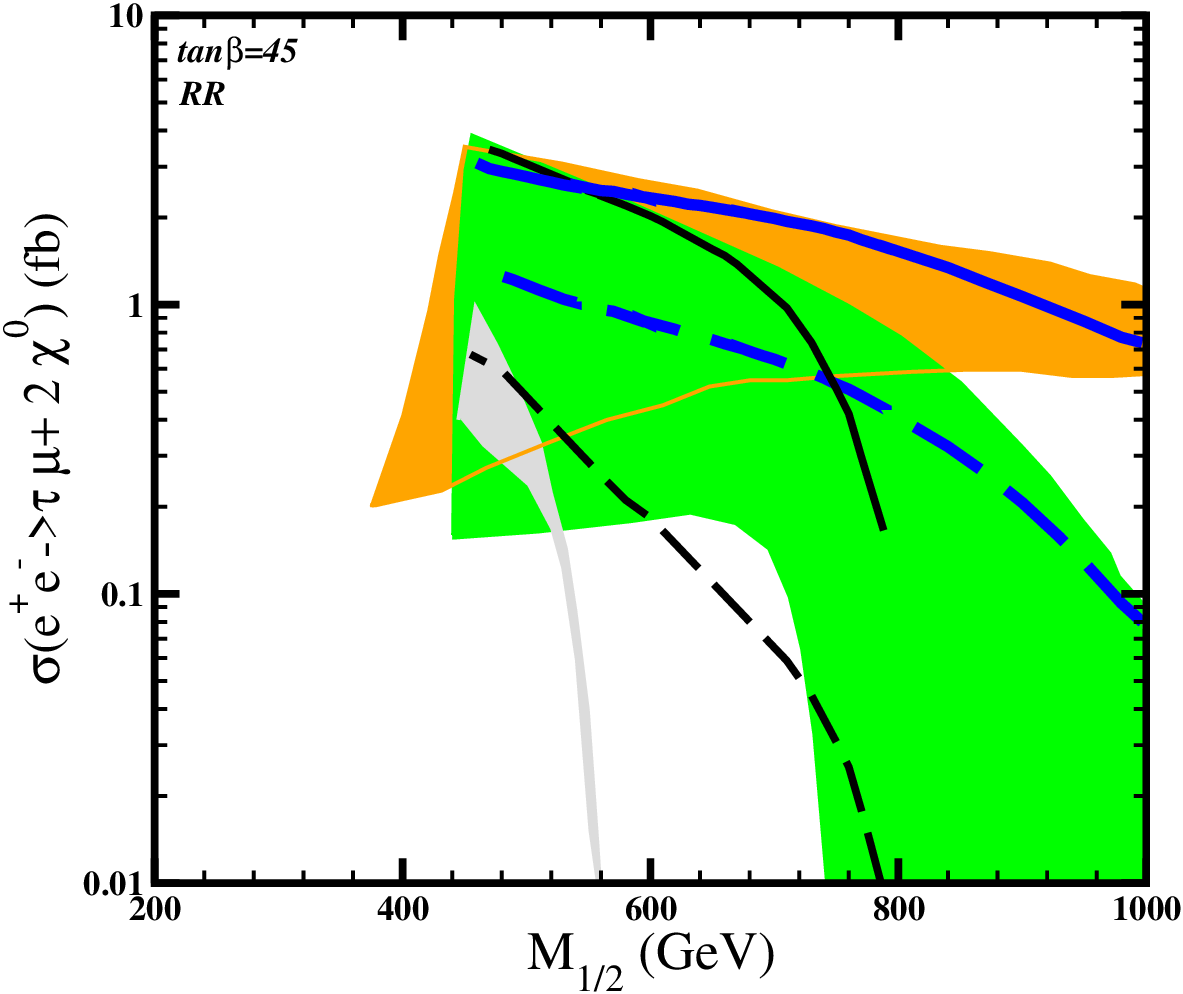} 
\end{center}
\caption{\small \it The shaded areas show the possible ranges of
$\sigma(e^+e^-\rightarrow \tau ^\pm\mu^\mp +2 \chi^0)$ for $m_0 <1000{\rm ~GeV}$, 
$\tan\beta=35$ (left panels)  and $\tan\beta=45$ (right panels), assuming $A_0=0$ in both cases,
with $\sqrt{s}$ fixed to $500{\rm ~GeV}$ (grey), $1000{\rm ~GeV}$ (green) 
and $2000{\rm ~GeV}$ (orange). In the upper panels, we assume the 
 left mixing parameters (\ref{eq:choiceL}), whereas in the lower panels we assume the 
right mixing (\ref{eq:choiceR}). The solid lines present the possible values for models along the center of the WMAP strips,
with  the lines corresponding to $\sqrt{s}$=500, 1000, 2000 ${\rm ~GeV}$ being progressively thicker.
The corresponding predictions for $\sigma(e^+e^-\rightarrow \tau ^\pm e^\mp +2 \chi^0)$ are shown by the dashed lines.
}
\label{fig:csm12_area}
\end{figure}

In Fig.~\ref{fig:csm12_area} we present  the maximum values of the cross sections 
in the areas allowed by all the current constraints. 
The shaded areas show the possible ranges of
$\sigma(e^+e^-\rightarrow \tau ^\pm\mu^\mp +2 \chi^0)$, and we
note that points along the WMAP strips (solid lines) generally have high values of the cross sections.
The dashed lines show the possible values of 
$\sigma(e^+e^-\rightarrow \tau ^\pm e^\mp +2 \chi^0)$ along the WMAP strips,
and we see that these cross sections may be of the 
same order of magnitude as those for $\mu-\tau$ pairs~%
\footnote{We do not 
display values for $\sigma(e^+e^-\rightarrow \mu ^\pm e^\mp +2 \chi^0)$, because this would
require a more precise analysis of mixings in the $LR$ sector.}.

\begin{figure}[!t]
\begin{center}
\includegraphics[scale=.40]{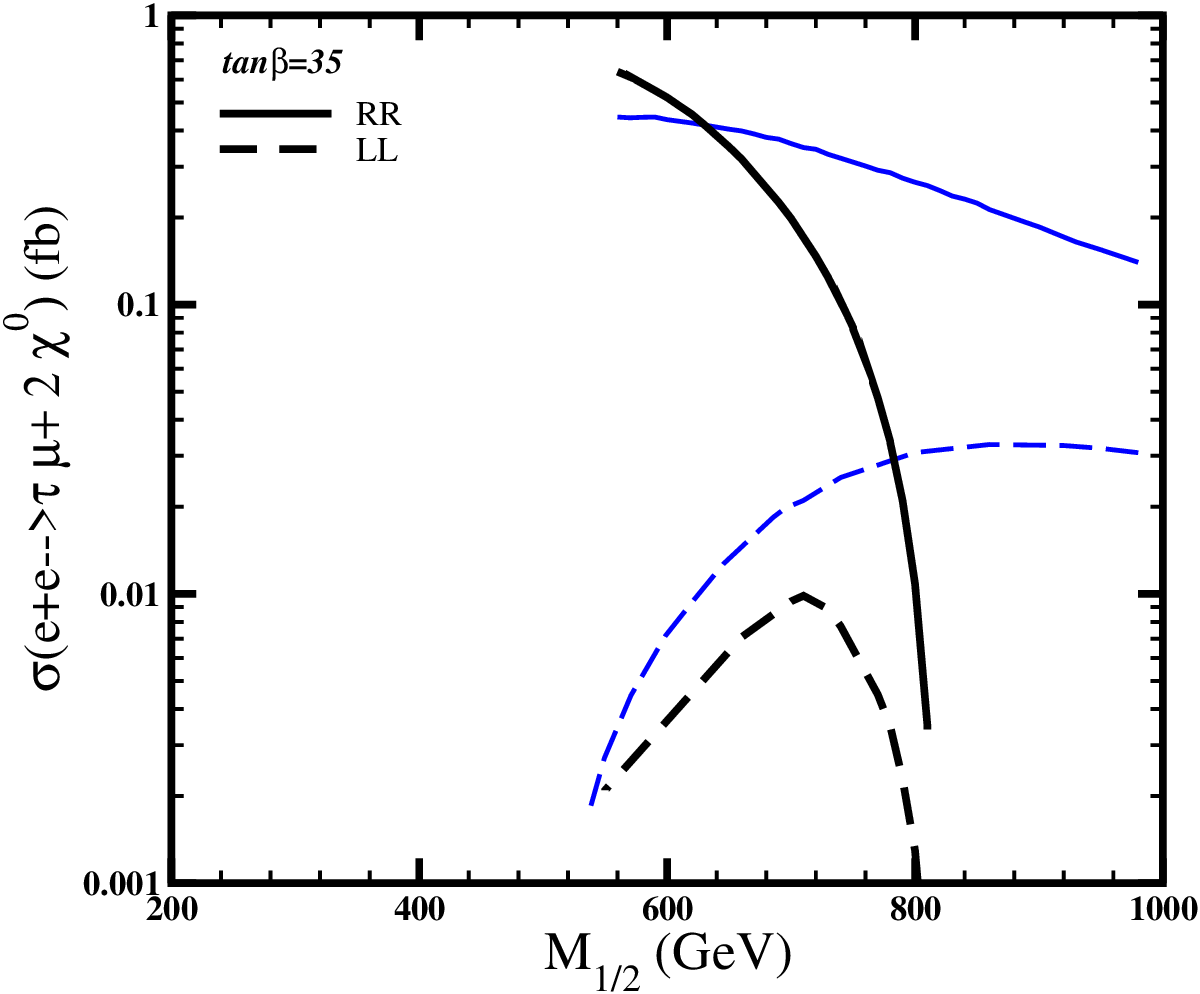}
\includegraphics[scale=.40]{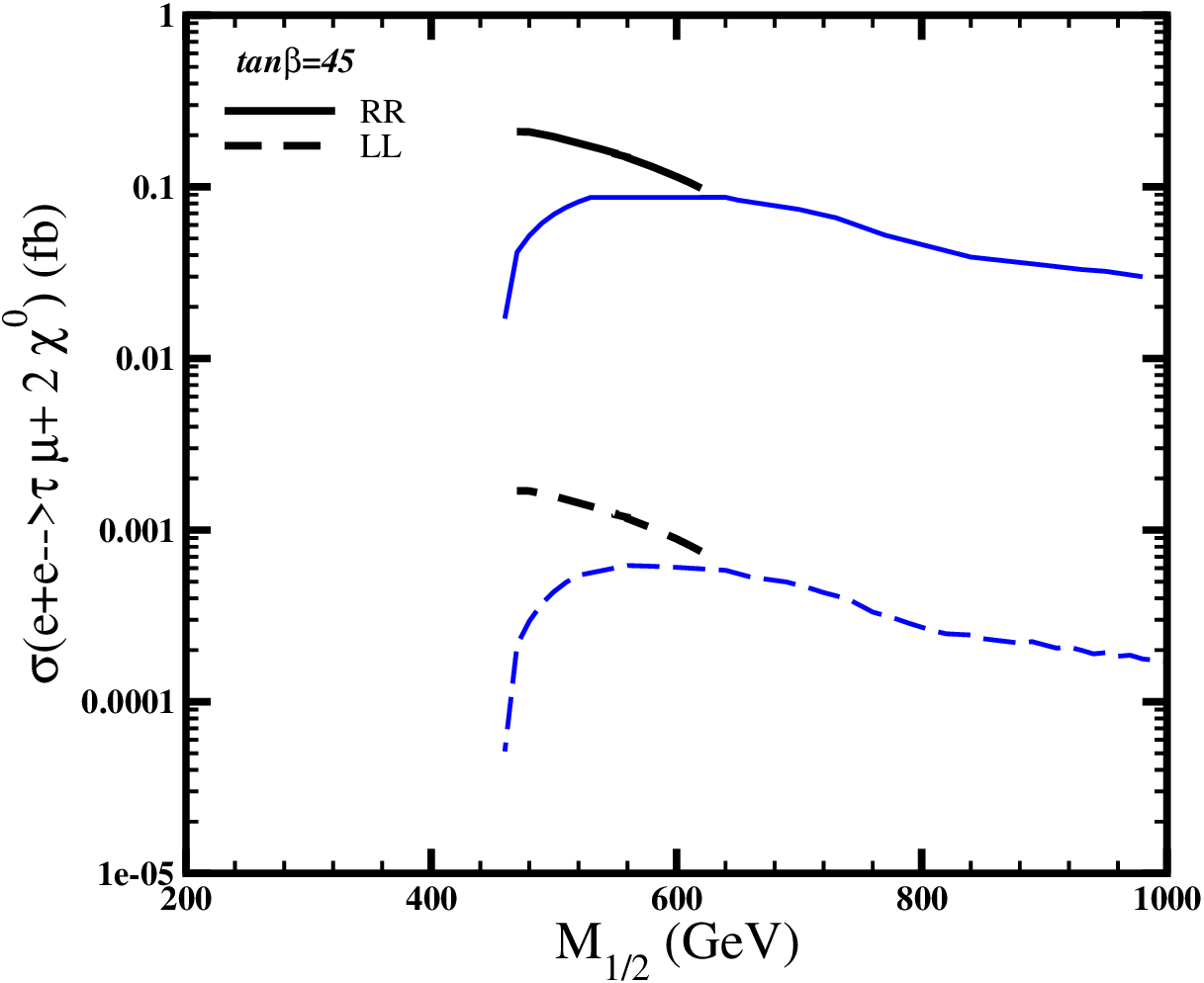}

\end{center}
\caption{\small \it Predictions from $\chi^0_2$ cascade decays (\ref{eq:single}) along the 
WMAP strips for $\sqrt{s}$= 1000 ${\rm ~GeV}$ (thick black lines) and 2000 ${\rm ~GeV}$ (thin blue lines). 
These channels are not accessible at $\sqrt{s}$=500 ${\rm ~GeV}$. The solid (dashed) lines 
correspond to mixing in the $RR$ ($LL$) sectors.
}
\label{fig:csm12_sp}
\end{figure}

In Fig.~\ref{fig:csm12_sp} we display the 
expectations for the production of LFV 
$\tau-\mu$ pairs in the cascade decays of eq.~(\ref{eq:single}), which are suppressed compared to
the pair production process of eq.~(\ref{eq:pair}). 
The production of LFV pairs from $\chi^0_2$ decays is analogous to the corresponding process at 
the LHC~\cite{CEGLR2}. We observe that phenomenologically interesting 
values on the $LL$ sector induce very low cross sections, whereas mixing in the $RR$ 
sector may be more observable, as it enhances the expected rates.

The SM background can be as large as several tens of fb. However, 
it has been shown in ref.~\cite{LFV-LC2} that  after imposing 
several cuts, it is possible to distinguish  LFV signals at the level 
 of 1~fb. 

The SUSY background  arises from flavor conserving channels like  
$e^+e^-\rightarrow \tilde{\nu}_i\bar{\tilde{\nu}}_i$,  
$e^+e^-\rightarrow \tilde{l}_i^+\tilde{l}_i^-$ and subsequent cascade decays 
that result in $\tau^\pm \mu^\mp + E\!\!\!\!/$ .  This 
background  is very dependent on the 
SUSY spectrum and is much smaller than the SM one. In our case, the branching ratios for $e^\pm \tau^\mp+ E\!\!\!\!/$ and 
$\mu^\pm \tau^\mp+ E\!\!\!\!/$ arising from the production of 
$\tilde{e}_L^+\tilde{e}_L^-$ and $\tilde{\mu}_L^+\tilde{\mu}_L^-$ are 
almost zero at the 
points with $\tan\beta=35$, and of about 10\% at the points $a45$ and $b45$.
Without considering any cut,  
this induces a background of less than 1~fb  in  $\tau^\pm \mu^\mp + E\!\!\!\!/$ 
at energies below 2 TeV.  At 2 TeV, the background 
for  $e^\pm \tau^\mp+ E\!\!\!\!/$ is bigger  due to the higher production of $\tilde{e}_L$ and $\tilde{\nu}$ in $e^\pm e^\mp$ collisions (while at $b45$  it remains of the order of 1~fb, it 
reaches a value of 12~fb at the point $a45$). Other channels like
$e^+e^-\rightarrow \tilde{\chi}_i^\pm\bar{\tilde{\chi}}^\mp_j$  and 
$e^+e^-\rightarrow \tilde{\chi}_1^+ e^-\bar{\tilde{\nu}_e}$ produce backgrounds 
 at the fb level even at 2 TeV.

\section{Conclusions}

Motivated by the challenging possibility of observable charged lepton flavour violation,
which would provide a new window on flavour physics that illuminates novel
aspects of supersymmetry,
we have explored possible signatures at the Linear Collider (LC), combining
phenomenological constraints with the cosmological considerations implied by WMAP.
We have found that both direct slepton pair production and sleptons produced in cascade decays 
may provide interesting signals
in the cosmologically-favoured region of the supersymmetric parametric
space. Moreover, the LC could provide additional insights beyond those obtainable from the LHC,
by virtue of its greater kinematic range for slepton production and its sensitivity to
$LL$ mixing as well as $RR$ mixing.

Within this framework, we found the following.

$\bullet$
The LC enhances significantly the prospects of detecting LFV for
heavy sparticle spectra, where flavour-violating rare decays and
conversions are significantly suppressed.  This allows probing an
entirely different range of the flavour-violating parameters.

$\bullet$
Unlike the LHC of ref.~\cite{CEGLR2}, at the LC we can have significant LFV within the CMSSM via $LL$ mixing.

$\bullet$
The decay $e^+e^-\rightarrow \tau ^\pm\mu^\mp +2 \chi^0$
provides an optimal search channel in this respect. Channels
where charginos play a significant role may require
a departure towards theories where gaugino or Higgs unification is
broken, in order to respect the conditions imposed by cosmology.

$\bullet$
Comparing direct slepton pair production with indirect slepton
production through the cascade decays of heavier sparticles,
we find that in the latter case, the cross sections induced by $LL$ mixing 
are approximately one order of magnitude lower than in the $RR$ mixing case. This
is consistent with the results previously obtained for the LHC in 
Ref.~\cite{CEGLR2}, and is due to the presence of additional
  channels in cascade pair production. These lead to an interesting
  scaling of the cross sections with the available energy, which is
  sensitive to slepton mixing parameters of the three generations.

It would be very interesting to further investigate the following issues.

$\bullet$
The cross sections expected in non-minimal extensions of the theory
might not only enhance channels that in the current scheme are
more suppressed, but could also enable a
comparison of the allowed range of mixing parameters in different
models.

$\bullet$
The expectations for models of massive neutrinos, in which
quantum corrections provide a significant source of LFV in the $LL$
channel, provide a potential link between LC observables and neutrino mass and mixing
parameters.

$\bullet$
Detailed simulations of signals and backgrounds are desirable.

Overall, it seems that the LC provides an optimal environment for the
study of LFV, whereas the LHC is limited to specific channels
that have significant backgrounds. The fact that the LC opens up
additional
possibilities may prove significant for making the link between
observable cross sections and flavour model building.


\vspace*{0.3 cm}
{\bf Acknowledgements} 
We would like to thank M. Cannoni and J. Rodr\'{\i}guez-Quintero for useful discussions. The work of J.E. is supported partly by the London
Centre for Terauniverse Studies (LCTS), using funding from the European
Research Council 
via the Advanced Investigator Grant 267352. S.L. thanks the CERN Theory Division for its kind hospitality.
The work of M.E.G. is supported by the
Spanish MCINN Consolider-Ingenio 2010 Programme under grant MultiDark CSD2009-00064, the project FPA2006-13825 and the project P07FQM02962 funded by ``Junta de Andalucia''.

\vskip 1. cm



\begin{thebibliography}{99}

\bibitem{skatm}
Y.~Fukuda {\it et al.}  [Super-Kamiokande Collaboration],
Phys.\ Rev.\ Lett.\ {\bf 81} (1998) 1562.

\bibitem{sksol}
Y.~Fukuda {\it et al.}  [Super-Kamiokande Collaboration],
Phys.\ Rev.\ Lett.\ {\bf 82} (1999) 1810;
Phys.\ Rev.\ Lett.\ {\bf 82} (1999) 2430;
Q.~R.~Ahmad {\it et al.}  [SNO Collaboration],
Phys.\ Rev.\ Lett.\  {\bf 87} (2001) 071301.

\bibitem{KamLand}
K. Eguchi et al., KamLAND Collaboration, Phys. Rev. Lett. {\bf 90} (2003) 021802; T.
Araki et al., KamLAND Collaboration, Phys. Rev. Lett. {\bf 94} (2005) 081801.

\bibitem{K2K}
M.H. Ahn et al., K2K Collaboration, Phys. Rev. Lett. {\bf 90} (2003) 041801.

\bibitem{MINOS}
D.G. Michael et al., MINOS colaboration, Phys. Rev. Lett. {\bf 97} (2006) 191801.

\bibitem{bm}
F.~Borzumati and A.~Masiero,
Phys.\ Rev.\ Lett.\ {\bf 57} (1986) 961.

\bibitem{rev}
Y. Kuno and  Y. Okada, 
Rev.\ Mod.\ Phys.\ {\bf 73} (2001) 151.

\bibitem{LFVhisano}
J.~Hisano, T.~Moroi, K.~Tobe and M.~Yamaguchi,
Phys.\ Rev.\ D {\bf 53} (1996) 2442.

\bibitem{LFVres}
J.~Hisano, D.~Nomura and T.~Yanagida,
Phys.\ Lett.\ B {\bf 437} (1998) 351;
W.~Buchm\"uller, D.~Delepine and F.~Vissani,
Phys.\ Lett.\ B {\bf 459} (1999) 171;
M.~E.~G\'omez, G.~K.~Leontaris, S.~Lola and J.~D.~Vergados,
Phys.\ Rev.\ D {\bf 59} (1999) 116009;
J.~R.~Ellis, M.~E.~G\'omez, G.~K.~Leontaris, S.~Lola and D.~V.~Nanopoulos,
Eur.\ Phys.\ J.\ C {\bf 14} (2000) 319;
W.~Buchm\"uller, D.~Delepine and L.~T.~Handoko,
Nucl.\ Phys.\ B {\bf 576} (2000) 445;
J.~L.~Feng, Y.~Nir and Y.~Shadmi,
Phys.\ Rev.\ D {\bf 61} (2000) 113005;
J.~Sato and K.~Tobe,
Phys.\ Rev.\ D {\bf 63} (2001) 116010;
J.~Hisano and K.~Tobe,
Phys.\ Lett.\ B {\bf 510} (2001) 197;
S.~Baek, T.~Goto, Y.~Okada and K.~Okumura,
Phys.\ Rev.\  D {\bf 64} (2001) 095001; 
S. Lavignac, I. Masina and C.A. Savoy, 
Phys.\ Lett.\  B {\bf 520}, 269 (2001);
D. Carvalho, J. Ellis, M. G\'omez and S. Lola,
Phys.\ Lett.\ B {\bf 515} (2001) 323;
T.~Blazek and S.~F.~King,
Phys. Lett. B {\bf 518} (2001) 109; 
M.~Cannoni, S.~Kolb and O.~Panella,
Phys.\ Rev.\  D {\bf 68}, 096002 (2003).

\bibitem{sleptonscemu}
N.~Arkani-Hamed, H.~Cheng, J.~L.~Feng and L.~J.~Hall,
Phys.\ Rev.\ Lett.\  {\bf 77} (1996) 1937;
Nucl.\ Phys.\ B {\bf 505} (1997) 3.

\bibitem{LFV-LC}
 J.~Hisano, M.~M.~Nojiri, Y.~Shimizu and M.~Tanaka,
  Phys.\ Rev.\  D {\bf 60}, 055008 (1999).

\bibitem{LC21}  M.~Guchait, J.~Kalinowski and P.~Roy,
  Eur.\ Phys.\ J.\  C {\bf 21}, 163 (2001).

\bibitem{LC2}
F.~Deppisch, J.~Kalinowski, H.~Pas, A.~Redelbach and R.~Ruckl,
  arXiv:hep-ph/0401243, work performed for the LHC study group.

\bibitem{LFV-LC2}
F. Deppisch, H. Pas, A. Redelbach, R. Ruckl, Y. Shimizu,
Phys. Rev. D69 (2004) 054014.

\bibitem{LHC1}
N.~V.~Krasnikov,
JETP Lett.\  {\bf 65} (1997) 148;
S.~I.~Bityukov and N.~V.~Krasnikov,
arXiv:hep-ph/9806504,
10th International Seminar on High-Energy Physics Suzdal, Russia;
K.~Agashe and M.~Graesser,
Phys.\ Rev.\ D {\bf 61} (2000) 075008.

\bibitem{LHC2}
J.~Hisano, R.~Kitano and M.~M.~Nojiri,
Phys.\ Rev.\ D {\bf 65} (2002) 116002;
I.~Hinchliffe and F. E.~Paige,
Phys.\ Rev.\ D {\bf 63} (2001) 115006;
D.Carvalho, J. Ellis, M. G\'omez, S. Lola, J.Romao,
Phys. Lett. B {\bf 618} (2005) 162.

\bibitem{CEGLR2}
  E.~Carqu\'in, J.~Ellis, M.~E.~G\'omez, S.~Lola and J.~Rodr\'iguez-Quintero,
JHEP 0905 (2009) 026. 

\bibitem{LEPH}
R.~Barate {\it et al.}  [ALEPH, DELPHI, L3, OPAL
                       Collaborations and LEP Working Group for Higgs
                       boson searches],
  Phys.\ Lett.\  B {\bf 565} (2003) 61.

  
\bibitem{LHC}
V.~Khachatryan {\it et al.}, CMS Collaboration,
  arXiv:1101.1628 [hep-ex];
 G.~Aad {\it et al.} [ATLAS Collaboration],
arXiv:1102.2357 [hep-ex];
G.~Aad {\it et al.} [ATLAS Collaboration],
  arXiv:1102.5290 [hep-ex].

\bibitem{MC6}
O.~Buchmueller {\it et al.},
  arXiv:1106.2529 [hep-ph];
  and references therein.
  
\bibitem{WMAP}
E.~Komatsu {\it et al.}  [WMAP Collaboration],
  Astrophys.\ J.\ Suppl.\  {\bf 192} (2011) 18.
  
\bibitem{JK-APP} 
J.~Kalinowski, 
Acta Phys.\ Polon.\ B {\bf 32} (2001) 3755.

\bibitem{pedro} 
M.E. G\'omez, S. Lola, P. Naranjo and P. Rodr\'iguez-Quintero, 
HEP 0904 (2009) 043. 

\bibitem{Martin:1993zk}
S.~P.~Martin and M.~T.~Vaughn, 
Phys.\ Rev.\  D {\bf 50}, 2282 (1994) 
[Erratum-ibid.\  D {\bf 78}, 039903 (2008)] 
[arXiv:hep-ph/9311340].

\bibitem{GNIS1}
M.~E.~Gomez, T.~Ibrahim, P.~Nath and S.~Skadhauge, 
Phys.\ Rev.\  D {\bf 72}, 095008 (2005) 
[arXiv:hep-ph/0506243].


\bibitem{mtop}
[Tevatron Electroweak Working Group and CDF Collaboration and D0 Collab], 
arXiv:0803.1683 [hep-ex].

\bibitem{GNIS0}
M.~E.~Gomez, T.~Ibrahim, P.~Nath and S.~Skadhauge, 
Phys.\ Rev.\ D {\bf 70}, 035014 (2004) 
[arXiv:hep-ph/0404025].

\bibitem{MicrOMEGAs}{MicrOMEGAs}
{\tt http://lapth.in2p3.fr/micromegas/}.

\bibitem{CalcHEP}
{\tt http://theory.sinp.msu.ru/~pukhov/calchep.html}.

\bibitem{suspect}
A.~Djouadi, J.~L.~Kneur and G.~Moultaka, 
Comput.\ Phys.\ Commun.\ {\bf 176}, 426 (2007) 
[arXiv:hep-ph/0211331].

\bibitem{Adam:2011ch}
  J.~Adam {\it et al.} [ MEG Collaboration ],
[arXiv:1107.5547 [hep-ex]].

\bibitem{PDG}
  K.~Nakamura {\it et al.} [Particle Data Group], J. Phys. G {\bf 37} (2010) 075021
and 2011 partial update for the 2012 edition: {\tt http://pdg.lbl.gov/}.

\bibitem{playground}
M.~Ciuchini, A.~Masiero, P.~Paradisi, L.~Silvestrini, S.~K.~Vempati and O.~Vives,
  Nucl.\ Phys.\  B {\bf 783} (2007) 112.


\end{thebibliography}
\end{document}